\@twosidetrue
\@twocolumntrue
\@mparswitchtrue
\def\ds@draft{\overfullrule 5pt}
\def\ds@twocolumn{\@twocolumntrue}
\def\ds@onecolumn{\@twocolumnfalse}
\newif\ifSFB@landscape
\def\ds@landscape{\SFB@landscapetrue}
\newif\ifSFB@galley
\def\ds@galley{\SFB@galleytrue}
\newif\ifSFB@referee
\def\ds@referee{%
 \SFB@refereetrue
 \@twocolumnfalse
}
\@options
\ifSFB@referee
 
\else
 
\fi
\if@twocolumn
  \def\@normalsize{\@setsize\normalsize{11pt}\ixpt\@ixpt
   \abovedisplayskip 6pt plus 2pt minus 2pt
   \belowdisplayskip \abovedisplayskip
   \abovedisplayshortskip 6pt plus 2pt
   \belowdisplayshortskip \abovedisplayshortskip
   \let\@listi\@listI}
 \else
 \ifSFB@referee
  \def\@normalsize{\@setsize\normalsize{14pt}\xiipt\@xiipt
   \abovedisplayskip 4pt plus 1pt minus 1pt
   \belowdisplayskip \abovedisplayskip
   \abovedisplayshortskip 4pt plus 1pt
   \belowdisplayshortskip \abovedisplayshortskip
   \let\@listi\@listI}
 \else
  \def\@normalsize{\@setsize\normalsize{12pt}\ixpt\@ixpt
   \abovedisplayskip 4pt plus 1pt minus 1pt
   \belowdisplayskip \abovedisplayskip
   \abovedisplayshortskip 4pt plus 1pt
   \belowdisplayshortskip \abovedisplayshortskip
   \let\@listi\@listI}
 \fi
\fi
\def\small{\@setsize\small{10pt}\viiipt\@viiipt
 \abovedisplayskip 4pt plus 1pt minus 1pt
 \belowdisplayskip \abovedisplayskip
 \abovedisplayshortskip 4pt plus 1pt
 \belowdisplayshortskip \abovedisplayshortskip
 \def\@listi{\leftmargin\leftmargini
  \topsep 2pt plus 1pt minus 1pt
  \parsep \z@
  \itemsep 2pt}}
\def\footnotesize{\@setsize\footnotesize{10pt}\viiipt\@viiipt
 \abovedisplayskip 4pt plus 1pt minus 1pt
 \belowdisplayskip \abovedisplayskip
 \abovedisplayshortskip 4pt plus 1pt
 \belowdisplayshortskip \abovedisplayshortskip
 \def\@listi{\leftmargin\leftmargini
  \topsep 2pt plus 1pt minus 1pt
  \parsep \z@
  \itemsep 2pt}}
\def\scriptsize{\@setsize\scriptsize{8pt}\viipt\@viipt}
\def\tiny{\@setsize\tiny{6pt}\vpt\@vpt}
\if@twocolumn
  \def\large{\@setsize\large{11pt}\xpt\@xpt}
 \else
  \def\large{\@setsize\large{12pt}\xpt\@xpt}
 \fi
\def\Large{\@setsize\Large{14pt}\xiipt\@xiipt}
\def\LARGE{\@setsize\LARGE{17pt}\xivpt\@xivpt}
\def\huge{\@setsize\huge{20pt}\xviipt\@xviipt}
\def\Huge{\@setsize\huge{25pt}\xxpt\@xxpt}
\normalsize

\if@twocolumn
\else
 \ifSFB@referee
  \oddsidemargin \z@  \evensidemargin \z@
 \else
 \fi
\fi
 \topmargin \z@
\newdimen\SFB@measure
\textwidth \SFB@measure
\ifSFB@landscape
 \textwidth \textheight
 \textheight \SFB@measure
\fi
\ifSFB@referee
\fi
\skip\footins 19.5pt plus 12pt minus 1pt

\parskip \z@ plus .1pt
\@beginparpenalty -\@lowpenalty
\@endparpenalty -\@lowpenalty
\@itempenalty -\@lowpenalty
\newcounter{part}
\newcounter {section}
\newcounter {subsection}[section]
\newcounter {subsubsection}[subsection]
\newcounter {paragraph}[subsubsection]
\newcounter {subparagraph}[paragraph]
\def\thepart          {\arabic{part}}
\def\thesection       {\arabic{section}}

\def\part{\par \addvspace{4ex}\@afterindentfalse
 \secdef\@part\@spart}
\def\@part[#1]#2{\ifnum \c@secnumdepth >\m@ne
  \refstepcounter{part}
  \addcontentsline{toc}{part}{Part \thepart: #1}
 \else \addcontentsline{toc}{part}{#1}
 \fi
 {\parindent 0pt \raggedright
  \ifnum \c@secnumdepth >\m@ne
   \large\rm PART
   \ifcase\thepart \or ONE \or TWO \or THREE \or FOUR \or FIVE
    \or SIX \or SEVEN \or EIGHT \or NINE \or TEN \else \fi
   \par \nobreak
  \fi
  \LARGE \rm #2 \markboth{}{}\par }
 \nobreak \vskip 3ex \@afterheading}
\def\@spart#1{{\parindent 0pt \raggedright
  \LARGE \rm #1\par}
 \nobreak \vskip 3ex \@afterheading}

\def\section{\@startsection {section}{1}{\z@}
 {-24pt plus -12pt minus -1pt}
 {6pt}
 {\SFB@hangraggedright\normalsize\bf}}
\def\subsection{\@startsection{subsection}{2}{\z@}
 {-18pt plus -9pt minus -1pt}
 {6pt}
 {\SFB@hangraggedright\large\bf}}
\def\subsubsection{\@startsection{subsubsection}{3}{\z@}
 {-18pt plus -9pt minus -1pt}
 {6pt}
 {\SFB@hangraggedright\normalsize\it}}
\def\paragraph{\@startsection{paragraph}{4}{\z@}
 {12pt plus 2.25pt minus 1pt}{-0.5em}{\normalsize\bf}}
\def\subparagraph{\@startsection{subparagraph}{5}{\parindent}
 {12pt plus 2.25pt minus 1pt}{-0.5em}{\normalsize\it}}
\def\SFB@hangraggedright{\rightskip\@flushglue \let\\=\newline}
\def\@sect#1#2#3#4#5#6[#7]#8{%
 \ifnum #2>\c@secnumdepth
  \def\@svsec{}%
 \else
  \refstepcounter{#1}
  \ifnum #2=\@ne
   \ifSFB@appendix \edef\@svsec{}%
             \else \edef\@svsec{\csname the#1\endcsname\hskip 1em}%
   \fi
  \else \edef\@svsec{\csname the#1\endcsname\hskip 1em}%
  \fi
 \fi
 \@tempskipa #5\relax
 \ifdim \@tempskipa>\z@
  \begingroup #6\relax
   \ifnum #2=\@ne
    \ifSFB@appendix
     \@hangfrom{\hskip #3\relax\@svsec}{\interlinepenalty \@M
      APPENDIX \csname the#1\endcsname:\hskip 0.5em\uppercase{#8}\par}%
    \else
     \@hangfrom{\hskip #3\relax\@svsec}{\interlinepenalty \@M
      \uppercase{#8}\par}%
    \fi
   \else
    \@hangfrom{\hskip #3\relax\@svsec}{\interlinepenalty \@M #8\par}%
   \fi
  \endgroup
  \csname #1mark\endcsname{#7}%
  \addcontentsline{toc}{#1}{\ifnum #2>\c@secnumdepth \else
   \protect\numberline{\csname the#1\endcsname}\fi #7}%
 \else
  \def\@svsechd{#6\hskip #3\@svsec \ifnum #2=\@ne\uppercase{#8}\else #8\fi
  \csname #1mark\endcsname{#7}
  \addcontentsline{toc}{#1}{\ifnum #2>\c@secnumdepth \else
   \protect\numberline{\csname the#1\endcsname}\fi#7}}%
 \fi
 \@xsect{#5}}

\newif\ifSFB@appendix
\def\appendix{\par
 \SFB@appendixtrue
 \setcounter{section}{0}
 \def\thesection{A\arabic{section}}
 \setcounter{equation}{0}
 \def\theequation{A\arabic{equation}}
 \setcounter{figure}{0}
 \def\thefigure{A\@arabic\c@figure}
 \setcounter{table}{0}
 \def\thetable{A\@arabic\c@table}
}

\newskip\@indentskip
\newskip\smallindent
\newskip\@footindent
\newskip\@leftskip
\@footindent=\smallindent
\@leftskip=\z@

\leftmargini   \@indentskip
\leftmargin\leftmargini
\labelwidth\leftmargini\advance\labelwidth-\labelsep
\def\makeRRlabel#1{\hss\llap{#1}}
\def\@listI{\leftmargin\leftmargini
 \parsep \z@
 \topsep 6pt plus 1pt minus 1pt
 \itemsep \z@ plus .1pt
}
\let\@listi\@listI
\@listi
\def\@listii{\leftmargin\leftmarginii
 \labelwidth\leftmarginii\advance\labelwidth-\labelsep
 \topsep 6pt plus 1pt minus 1pt
 \parsep \z@
 \itemsep \z@ plus .1pt
}
\def\@listiii{\leftmargin\leftmarginiii
 \labelwidth\leftmarginiii\advance\labelwidth-\labelsep
 \topsep 6pt plus 1pt minus 1pt
 \parsep \z@
 \partopsep \z@
 \itemsep \topsep
}
\def\@listiv{\leftmargin\leftmarginiv
 \labelwidth\leftmarginiv\advance\labelwidth-\labelsep
}
\def\@listv{\leftmargin\leftmarginv
 \labelwidth\leftmarginv\advance\labelwidth-\labelsep
}
\def\@listvi{\leftmargin\leftmarginvi
 \labelwidth\leftmarginvi\advance\labelwidth-\labelsep
}
\def\itemize{\ifnum \@itemdepth >3 \@toodeep
  \else \advance\@itemdepth \@ne
   \edef\@itemitem{labelitem\romannumeral\the\@itemdepth}%
   \list{\csname\@itemitem\endcsname}%
    {\let\makelabel\makeRRlabel}%
  \fi}

\def\enumerate{\ifnum \@enumdepth >3 \@toodeep \else
  \advance\@enumdepth \@ne
  \edef\@enumctr{enum\romannumeral\the\@enumdepth}%
 \fi
 \@ifnextchar [{\@enumeratetwo}{\@enumerateone}%
}
\def\@enumeratetwo[#1]{%
 \list{\csname label\@enumctr\endcsname}%
  {\settowidth\labelwidth{[#1]}
   \leftmargin\labelwidth \advance\leftmargin\labelsep
   \usecounter{\@enumctr}
   \let\makelabel\makeRRlabel}
}
\def\@enumerateone{%
 \list{\csname label\@enumctr\endcsname}%
  {\usecounter{\@enumctr}
   \let\makelabel\makeRRlabel}}

\def\theenumi{(\roman{enumi})}

\def\theenumii{(\alph{enumii})}
\def\p@enumii{\theenumi}

\def\theenumiii{(\arabic{enumiii})}
\def\p@enumiii{\theenumi(\theenumii)}

\def\p@enumiv{\p@enumiii\theenumiii}
\def\description{\list{}{\labelwidth\z@ \itemindent-\leftmargin
  \leftmargin 1em
  \itemindent-1em
}}

\def\verse{\let\\=\@centercr
 \list{}{\itemsep\z@
  \itemindent -\@indentskip
  \listparindent \itemindent
  \rightmargin\leftmargin
  \advance\leftmargin \@indentskip}\item[]}

\def\quotation{\list{}{\listparindent \smallindent
  \leftmargin\z@\rightmargin\leftmargin
  \parsep 0pt plus 1pt}\item[]\small}

\def\quote{\list{}{\leftmargin\z@\rightmargin\leftmargin}\item[]\small}

\def\@begintheorem#1#2{\rm \trivlist \item[\hskip \labelsep{\bf #1\ #2.}]}
\def\@opargbegintheorem#1#2#3{\rm \trivlist
  \item[\hskip \labelsep{\bf #1\ #2.\ (#3)}]}
\def\@endtheorem{\endtrivlist}
\@namedef{proof*}{\rm \trivlist \item[\hskip \labelsep{\it Proof.}]}
\@namedef{endproof*}{\endtrivlist}

\def\titlepage{\@restonecolfalse\if@twocolumn\@restonecoltrue\onecolumn
  \else \newpage \fi \thispagestyle{empty}\c@page\z@}
\def\endtitlepage{\if@restonecol\twocolumn \else \newpage \fi}

\def\tabular{\def\@halignto{}
 \def\hline{\noalign{\ifnum0=`}\fi
  \vskip 3pt
  \hrule \@height \arrayrulewidth
  \vskip 3pt
  \futurelet \@tempa\@xhline}
 \def\fullhline{\noalign{\ifnum0=`}\fi
  \vskip 3pt
  \hrule \@height \arrayrulewidth
  \vskip 3pt
  \futurelet \@tempa\@xhline}
 \def\@xhline{\ifx\@tempa\hline
   \vskip -6pt
   \vskip \doublerulesep
  \fi
  \ifnum0=`{\fi}}
  \def\@arrayrule{\@addtopreamble{\hskip -.5\arrayrulewidth
                                  \hskip .5\arrayrulewidth}}
\@tabular
}
\tabbingsep \labelsep

\skip\@mpfootins = \skip\footins

\fboxrule = \arrayrulewidth

\def\maketitle{\par
 \begingroup
  \if@twocolumn
   \twocolumn[\vspace*{17pt}\@maketitle]
  \else
   \newpage
   \global\@topnum\z@
   \@maketitle
  \fi
  \thispagestyle{titlepage}
 \endgroup
 \let\maketitle\relax
 \let\@maketitle\relax
 \gdef\@author{}
 \gdef\@title{}
 \let\thanks\relax
}
\def\and{\end{author@tabular}\vskip 6pt\par
 \begin{author@tabular}[t]{@{}l@{}}}
\def\@maketitle{\newpage
 \vspace*{7pt}
 {\raggedright \sloppy
  {\huge \bf \@title \par}
  \vskip 23pt
  {\LARGE
   \begin{author@tabular}[t]{@{}l@{}}\@author
   \end{author@tabular}\par}
  \vskip 22pt
 }
 \par\noindent
 {\small \@date \par}
 \vskip 22pt
}
\def\abstract{\if@twocolumn
  \start@SFBbox\@abstract
 \else
  \@abstract
 \fi}
\def\endabstract{\if@twocolumn
   \endlist\finish@SFBbox
 \else
  \endlist
 \fi}
\def\@abstract{\list{}{\leftmargin 10.5pc\rightmargin\z@
  \parsep 0pt plus 1pt}\item[]\normalsize{\bf ABSTRACT}\\\large} 
\newif\ifSFB@keywords
\def\keywords{\if@twocolumn
  \start@SFBbox\@keywords
 \else
  \@keywords
 \fi
}
\def\@keywords{\list{}{\leftmargin 10.5pc\rightmargin\z@
  \parsep 0pt plus 1pt}\item[]\large{\bf Key words: }}
\def\endkeywords{\if@twocolumn
  \endlist\addvspace{37pt}\finish@SFBbox
 \else
  \endlist
 \fi
 \@thanks
 \gdef\@thanks{}
 \SFB@keywordstrue
}
\def\nokeywords{\ifSFB@keywords\else
 \if@twocolumn \start@SFBbox\addvspace{37pt}\finish@SFBbox \fi
 \@thanks
 \gdef\@thanks{}\fi
}

\def\author@tabular{\def\@halignto{}\@authortable}
\let\endauthor@tabular=\endtabular
\def\author@tabcrone{{\ifnum0=`}\fi\@xtabularcr[-7pt]\small\it
 \let\\=\author@tabcrtwo\ignorespaces}
\def\author@tabcrtwo{{\ifnum0=`}\fi\@xtabularcr[-7pt]\small\it
 \let\\=\author@tabcrtwo\ignorespaces}
\def\@authortable{\leavevmode \hbox \bgroup $\let\@acol\@tabacol
 \let\@classz\@tabclassz \let\@classiv\@tabclassiv
 \let\\=\author@tabcrone \ignorespaces \@tabarray}

\def\start@SFBbox{\@next\@currbox\@freelist{}{}%
 \global\setbox\@currbox
 \vbox\bgroup
  \hsize \textwidth
  \@parboxrestore
}
\def\finish@SFBbox{\par\vskip -\dbltextfloatsep
  \egroup
  \global\count\@currbox\tw@
  \global\@dbltopnum\@ne
  \global\@dbltoproom\maxdimen\@addtodblcol
  \global\vsize\@colht
  \global\@colroom\@colht
}

\mark{{}{}}
\gdef\@author{\mbox{}}
\def\author{\@ifnextchar [{\@authortwo}{\@authorone}}
\def\@authortwo[#1]#2{\gdef\@author{#2}\gdef\@shortauthor{#1}}
\def\@authorone#1{\gdef\@author{#1}\gdef\@shortauthor{#1}}
\gdef\@shortauthor{}
\gdef\@title{\mbox{}}
\def\title{\@ifnextchar [{\@titletwo}{\@titleone}}
\def\@titletwo[#1]#2{\gdef\@title{#2}\gdef\@shorttitle{#1}}
\def\@titleone#1{\gdef\@title{#1}\gdef\@shorttitle{#1}}
\gdef\@shorttitle{}
\def\volume#1{\gdef\@volume{#1}}
\gdef\@volume{000}
\def\microfiche#1{\gdef\@microfiche{#1}}
\gdef\@microfiche{}
\def\pagerange#1{\gdef\@pagerange{#1}}
\gdef\@pagerange{000--000}
\def\journal#1{\gdef\@journal{#1}}
\gdef\@journal{{Mon.\ Not.\ R.\ Astron.\ Soc.} {\bf \@volume}, \@pagerange\
  (\number\year) \@microfiche}
\def\ps@headings{\let\@mkboth\markboth
 \def\@oddhead{\Large \hfill \it \@shorttitle \hspace{1.5em}\rm \thepage}
 \def\@oddfoot{}
 \def\@evenhead{\Large \thepage \hspace{1.5em}\it \@shortauthor \hfill}
 \def\@evenfoot{}
 \def\sectionmark##1{\markboth{##1}{}}
 \def\subsectionmark##1{\markright{##1}}}
\def\ps@myheadings{\let\@mkboth\@gobbletwo
 \def\@oddhead{\Large \it \rightmark \hfill \rm \thepage}
 \def\@oddfoot{}
 \def\@evenhead{\Large \it \leftmark \hfill \rm \thepage}
 \def\@evenfoot{}
 \def\sectionmark##1{}
 \def\subsectionmark##1{}}
\def\ps@titlepage{\let\@mkboth\@gobbletwo
 \def\@oddhead{\footnotesize\@journal\hfill}
 \def\@oddfoot{}
 \def\@evenhead{\footnotesize\@journal\hfill}
 \def\@evenfoot{}
 \def\sectionmark##1{}
 \def\subsectionmark##1{}}

\def\@pnumwidth{1.55em}
\def\@tocrmarg {2.55em}
\def\@dotsep{4.5}
\def\@undottedtocline#1#2#3#4#5{\ifnum #1>\c@tocdepth
 \else
  \vskip \z@ plus .2pt
  {\hangindent #2\relax
   \rightskip \@tocrmarg \parfillskip -\rightskip
   \parindent #2\relax \@afterindenttrue
   \interlinepenalty\@M \leavevmode
   \@tempdima #3\relax #4\nobreak \hfill \nobreak
   \hbox to\@pnumwidth{\hfil\rm \ }\par}\fi}
\def\tableofcontents{\@restonecolfalse
 \if@twocolumn\@restonecoltrue\onecolumn\fi
 \section*{CONTENTS} \@starttoc{toc}
 \if@restonecol\twocolumn\fi \par\vspace{12pt}}
\def\l@part#1#2{\addpenalty{-\@highpenalty}
 \addvspace{2.25em plus 1pt}
 \begingroup
  \parindent \z@ \rightskip \@pnumwidth
  \parfillskip -\@pnumwidth
  {\normalsize\rm
   \leavevmode \hspace*{3pc}
   #1\hfil \hbox to\@pnumwidth{\hss \ }}\par
   \nobreak \global\@nobreaktrue
   \everypar{\global\@nobreakfalse\everypar{}}\endgroup}
\def\l@section#1#2{\addpenalty{\@secpenalty}
 \@tempdima 1.5em
 \begingroup
  \parindent \z@ \rightskip \@pnumwidth
  \parfillskip -\@pnumwidth \rm \leavevmode
  \advance\leftskip\@tempdima \hskip -\leftskip
  #1\nobreak\hfil \nobreak\hbox to\@pnumwidth{\hss \ }\par
 \endgroup}
\def\l@subsection{\@undottedtocline{2}{1.5em}{2.3em}}
\def\l@subsubsection{\@undottedtocline{3}{3.8em}{3.2em}}
\def\l@paragraph{\@undottedtocline{4}{7.0em}{4.1em}}
\def\l@subparagraph{\@undottedtocline{5}{10em}{5em}}
\def\listoffigures{\@restonecolfalse
 \if@twocolumn\@restonecoltrue\onecolumn\fi
 \section*{LIST OF FIGURES\@mkboth{LIST OF FIGURES}{LIST OF FIGURES}}
 \@starttoc{lof} \if@restonecol\twocolumn\fi}
\def\l@figure{\@undottedtocline{1}{1.5em}{2.3em}}
\def\listoftables{\@restonecolfalse
 \if@twocolumn\@restonecoltrue\onecolumn\fi
 \section*{LIST OF TABLES\@mkboth{LIST OF TABLES}{LIST OF TABLES}}
 \@starttoc{lot} \if@restonecol\twocolumn\fi}
\let\l@table\l@figure

\def\thebibliography#1{\section*{REFERENCES}
 \addcontentsline{toc}{section}{REFERENCES}
 \list{}{\labelwidth\z@
         \leftmargin 1.5em
	 \itemsep \z@
	 \itemindent-\leftmargin}
 \small\raggedright
 \parindent\z@
 \parskip\z@ plus .1pt\relax
 \def\newblock{\hskip .11em plus .33em minus .07em}
 \sloppy\clubpenalty4000\widowpenalty4000
 \sfcode`\.=1000\relax
}

\def\@biblabel#1{\hspace*{\labelsep}[#1]}

\newif\if@restonecol
\def\theindex{\section*{INDEX}
 \addcontentsline{toc}{section}{INDEX}
 \footnotesize \parindent\z@ \parskip\z@ plus .1pt\relax
 \let\item\@idxitem}
\def\@idxitem{\par\hangindent 1em}

\def\endtheindex{\if@restonecol\onecolumn\else\clearpage\fi}

\def\footnoterule{\kern-3\p@ \hrule width 12pc height \z@ \kern 3\p@}

\def\@fnsymbol#1{\ifcase#1\or \mbox{$\star$}\or \dagger\or \ddagger\or
   \S \or \P \or \|\or **\or \dagger\dagger
   \or \ddagger\ddagger\or \S\S\or \P\P\or \|\|\else ***
   \fi\relax}

\long\def\@makefntext#1{\parindent 1em\noindent
  $^{\@thefnmark}$\hspace{4pt}#1}

\newcounter{table}
\def\thetable{\@arabic\c@table}
\def\fps@table{tbp}
\def\ftype@table{1}
\def\ext@table{lot}
\def\fnum@table{Table \thetable}
\def\table{\let\@makecaption=\SFB@maketablecaption\@float{table}}
\let\endtable\end@float
\@namedef{table*}{\let\@makecaption=\SFB@maketablecaption\@dblfloat{table}}
\@namedef{endtable*}{\end@dblfloat}

\newcounter{figure}
\def\thefigure{\@arabic\c@figure}
\def\fps@figure{tbp}
\def\ftype@figure{2}
\def\ext@figure{lof}
\def\fnum@figure{Figure \thefigure}
\def\figure{\let\@makecaption=\SFB@makefigurecaption\@float{figure}}
\let\endfigure\end@float
\@namedef{figure*}{\let\@makecaption=\SFB@makefigurecaption\@dblfloat{figure}}
\@namedef{endfigure*}{\end@dblfloat}

\long\def\SFB@makefigurecaption#1#2{\vskip 6pt
 \setbox\@tempboxa\hbox{\small{\bf #1.} #2}
 \ifdim \wd\@tempboxa >\hsize
  \small{\bf #1.} #2\par
 \else
  \hbox to\hsize{\hfil\box\@tempboxa\hfil}
 \fi
 \vskip 6pt
}
\long\def\SFB@maketablecaption#1#2{\vskip 6pt
 \setbox\@tempboxa\hbox{\small{\bf #1.} #2}
 \ifdim \wd\@tempboxa >\hsize
  \small{\bf #1.} #2\par
 \else
  \hbox to\hsize{\box\@tempboxa\hfill}
 \fi
 \vskip 6pt
}

\def\caption{\@ifstar{\SFB@caption\@captype}%
 {\refstepcounter\@captype \@dblarg{\@caption\@captype}}%
}
\long\def\SFB@caption#1#2{
 \begingroup
  \@parboxrestore
  \normalsize
  \@makecaption{\csname fnum@#1\endcsname}{\ignorespaces #2}\par
 \endgroup}

\def\@cite#1#2{(#1\if@tempswa , #2\fi)}
\def\@biblabel#1{}

\newlength{\bibhang}
\def\@citex[#1]#2{\if@filesw\immediate\write\@auxout{\string\citation{#2}}\fi
  \def\@citea{}\@cite{\@for\@citeb:=#2\do
    {\@citea\def\@citea{; }\@ifundefined
       {b@\@citeb}{{\bf ?}\@warning
       {Citation `\@citeb' on page \thepage \space undefined}}%
{\csname b@\@citeb\endcsname}}}{#1}}
\let\@internalcite\cite
\def\cite{\def\citename##1{##1}\@internalcite}
\def\shortcite{\def\citename##1{}\@internalcite}

\def\[{\relax\ifmmode\@badmath\else\begin{trivlist}\item[]\leavevmode
  \hbox to\linewidth\bgroup$
  \displaystyle
  \hskip\mathindent\bgroup\fi}

\def\]{\relax\ifmmode \egroup $\hfil
       \egroup \end{trivlist}\else \@badmath \fi}

\def\equation{\refstepcounter{equation}\trivlist \item[]\leavevmode
  \hbox to\linewidth\bgroup $
  \displaystyle
\hskip\mathindent}

\def\endequation{$\hfil
           \displaywidth\linewidth\@eqnnum\egroup \endtrivlist}

\def\eqnarray{\stepcounter{equation}\let\@currentlabel=\theequation
\global\@eqnswtrue
\global\@eqcnt\z@\tabskip\mathindent\let\\=\@eqncr
\abovedisplayskip\topsep\ifvmode\advance\abovedisplayskip\partopsep\fi
\belowdisplayskip\abovedisplayskip
\belowdisplayshortskip\abovedisplayskip
\abovedisplayshortskip\abovedisplayskip
$$\halign
to \linewidth\bgroup\@eqnsel\hskip\@centering$\displaystyle\tabskip\z@
  {##}$&\global\@eqcnt\@ne \hskip 2\arraycolsep \hfil${##}$\hfil
  &\global\@eqcnt\tw@ \hskip 2\arraycolsep $\displaystyle{##}$\hfil
   \tabskip\@centering&\llap{##}\tabskip\z@\cr}

\def\endeqnarray{\@@eqncr\egroup
 \global\advance\c@equation\m@ne$$\global\@ignoretrue}

\newdimen\mathindent
\mathindent = \z@

\def\today{\number\day\ \ifcase\month\or
  January\or February\or March\or April\or May\or June\or
  July\or August\or September\or October\or November\or December
 \fi \ \number\year}

\ps@headings
\ifSFB@galley
 \ps@empty
\fi
\ifSFB@referee
\fi
\if@twocolumn
 \twocolumn
\else
 \onecolumn
\fi
\documentstyle[referee,seceqn]{mn}

\title[Infrared searches for brown dwarfs]{Infrared searches for dark matter in
       the form of brown dwarfs}

\author[E. J. Kerins \, \& \, B. J. Carr]
       {E. J. Kerins \, \& \, B. J. Carr \\
Astronomy Unit \\
School of Mathematical Sciences \\
Queen Mary \& Westfield College \\
Mile End Road \\
London, E1 4NS.}

\begin{document}

\maketitle

\newcommand{\be}{\begin{equation}}
\newcommand{\ee}{\end{equation}}
\newcommand{\iras}{{\em IRAS}\/ }
\newcommand{\iso}{{\em ISO}\/ }
\newcommand{\sirtf}{{\em SIRTF}\/ }
\newcommand{\cobe}{{\em COBE}\/ }
\newcommand{\dis}{\langle d(m) \rangle}
\newcommand{\disdel}{\langle d \rangle}
\newcommand{\cdis}{\langle D \rangle}
\newcommand{\mc}{M_{6}}
\newcommand{\mbd}{m_{0.01}}
\newcommand{\fre}{\nu_{13}}
\newcommand{\tbd}{\tau_{*}}
\newcommand{\k}{\kappa_{0.01}}
\newcommand{\tu}{\tau}
\newcommand{\ts}{\textstyle}
\newcommand{\ls}{\;\raisebox{-.8ex}{$\buildrel{\textstyle<}\over\sim$}\;}
\newcommand{\gs}{\;\raisebox{-.8ex}{$\buildrel{\textstyle>}\over\sim$}\;}

\begin{abstract}
Brown dwarfs, stars with insufficient mass to burn hydrogen, could contribute
to the dark matter in the Galactic disk, galactic halos or even a background
critical density. We consider the detectability of such brown dwarfs in various
scenarios, extending previous work by allowing for the possibility  that they
may have an extended mass spectrum or be clumped into dark clusters. We
investigate the constraints placed on such scenarios by the \iras survey.
Whilst an extrapolation of the mass function of visible disk stars makes it
unlikely that brown dwarfs comprise all of the proposed disk dark matter, \iras
does not exclude brown dwarfs providing the dark matter in our own halo or a
cosmological background. Neither does it improve on existing dynamical
constraints on the mass and radius of brown dwarf clusters in our halo. Future
satellites such as \iso and \sirtf will either detect brown dwarfs or brown
dwarf clusters or else severely constrain their contribution to the dark
matter.
\end{abstract}

\begin{keywords}
   Cosmology -- dark matter: Brown Dwarfs -- infrared detection.
\end{keywords}

\section{Introduction}

While ordinary visible material has a density $\Omega_{v}\approx 0.003$ in
units of the critical density \cite{persic92}, there is evidence for a much
larger density of invisible material (Faber \& Gallagher 1979; Trimble 1987;
Turner 1991). Dark matter has been proposed in four contexts: there may be {\em
local} dark matter associated with the Galactic disk ($\Omega \sim 0.001$),
dark
matter in galactic {\em halos} ($\Omega \sim 0.02-0.2$, depending on the
typical halo radius), dark matter in {\em clusters} of galaxies ($\Omega \sim
0.1-0.3$) and, if one accepts the inflationary scenario \cite{guth81}, {\em
background} dark matter in order to make the total density parameter unity. The
evidence for local and background dark matter is controversial but the
existence of halo and cluster dark matter seems unavoidable. The halo dark
matter may or may not be the same as the cluster dark matter. The values
indicated for $\Omega$ are very approximate but suffice to indicate that the
disk dark matter problem is distinct from the other ones and that the halo dark
matter may or may not be the same as the cluster dark matter. Despite its
subcritical density, the cluster dark matter could also be the same as the
background dark matter in biassed theories of galaxy formation.

\subsection{Evidence for baryonic dark matter}

Dark matter candidates may be classed as {\em non-baryonic} or {\em baryonic}.
In the former case the dark objects are relict particles from the very early
universe. Baryonic dark matter, the focus of this paper, is usually connected
with the formation of an early generation of stars, sometimes termed
``Population III'' objects (Rees 1978; Carr, Bond \& Arnett 1984). Such stars
could either be pregalactic, in which case they would have formed out of the
background gas around $10^{7} - 10^{9}$ years after the Big Bang and be
distributed throughout the universe, or protogalactic, in which case they would
form during the first phases of galaxy formation and be confined to halos.

The most important evidence for baryonic dark matter come from cosmological
nucleosynthesis considerations. Calculations by Walker et al.
\shortcite{walker91} show that the primordial abundances of the light elements
are explained if the baryon density parameter $\Omega_{B}$ lies in the range
$0.01\leq \Omega_{B} h^{2}\leq 0.015$ (where $h$ is the Hubble constant in
units of 100 km s$^{-1}$Mpc$^{-1}$). If $h=1$, as some people propose, then
$\Omega_{B} \geq 0.01$ and so at least two-thirds of the baryons must be dark.
In this case, dark baryons could not account for the halo dark matter but they
could still provide the local disk dark matter. However, with such a large
Hubble constant the age of the universe would be less than the age of the
oldest globular clusters unless there was a non-zero cosmological constant. For
values of $h$ less than 0.75, as many people prefer, $\Omega_{B} > 0.2$ and so
baryons can provide most or all, for modest halo radii,  of the halo dark
matter. Whatever the value of $h$, the missing baryons could still be contained
in a hot intergalactic medium. However, there is no direct evidence for this
and the \cobe limits on the spectral distortion of the microwave background
\cite{mather90}, together with the Gunn-Peterson constraint, implies that its
temperature would need to be in a narrow range \cite{barcons91}.

Dark baryons could only explain the dark matter in clusters or in the
background if it is possible to weaken the nucleosynthesis bounds by
invoking inhomogeneous nucleosynthesis (Applegate, Hogan \& Scherrer
1987; Alcock, Fuller \& Mathews 1987; Malaney \& Fowler 1988). The idea is that
a first order phase transition during the quark-hadron era would give rise to
fluctuations in the baryon density. Neutrons would then diffuse from the
overdense regions, due to their smaller cross-section, leading to variations in
the neutron-to-proton ratio. In the most optimistic case, it may be possible to
have $\Omega_{B} = 0.088 h^{-2}$ \cite{mathews93}, in which case all of the
dark matter in galaxies and clusters could be baryonic for low values of $h$.
Possible support for a high baryon density comes from {\em ROSAT}\/
observations of the Coma cluster, where the observed gas fraction at the Abell
radius is substantially higher than predicted for an $\Omega = 1, \; \Omega_{B}
\leq 0.015 h^{-2}$ universe \cite{frenk93}. However, the recent \cobe microwave
background anisotropy results \cite{smoot92} suggest that a baryon-dominated
universe is inconsistent with scenarios in which galaxies originate from
gravitational instability \cite{efstathiou92}, although models with ``$n = 0$''
isocurvature fluctuations might just be viable.

\subsection{Baryonic dark matter candidates}

Carr \shortcite{carr90} has reviewed the various constraints on halo baryonic
candidates. He concludes that the halo objects, if baryonic, must either be the
black hole remnants of very massive stars or very low mass stars. Intermediate
mass stars could produce white dwarf or neutron star remnants. However, {\em
white dwarfs} could probably only be relevant to explaining disk dark matter
else they would overproduce helium. Ryu, Olive \& Silk \shortcite{ryu90} have
argued that they could also comprise the halo but only if the mass function is
strongly peaked around $5 \; M_{\odot}$. Isolated {\em neutron stars} are
probably excluded from explaining any of the dark matter problems because their
precursors would overproduce heavy elements, although Salpeter \& Wasserman
\shortcite{salpeter93} have suggested that this could be avoided if the neutron
stars were in {\em clusters} whose gravitational potential and high gas density
traps any heavy elements produced. This possibility may be incompatible with
background light constraints \cite{kerins93}.

The {\em black hole} hypothesis was examined in detail by Carr, Bond \& Arnett
\shortcite{carr84}. They argued that the holes would most plausibly be the
remnants of ``Very Massive Objects'' larger than $200 \; M_{\odot}$, which
undergo gravitational collapse with no metal ejection as a result of an
instability during their oxygen-burning phase. The black hole remnants of
``Supermassive Objects'', stars larger than $10^{5} \; M_{\odot}$ which
collapse due to relativistic instabilities without undergoing any nuclear
burning at all, are probably excluded by dynamical constraints (Lacey \&
Ostriker 1985, Carr \& Sakellariadou 1993). Unfortunately, the VMO scenario now
seems less plausible because the {\em FIRAS}\/ results \cite{mather93} show no
signs of the infrared background radiation which would be an inevitable
consequence of the VMOs main-sequence phase \cite{bond91}. VMOs are not yet
definitely excluded, because there is uncertainty as to whether the radiation
would have been reprocessed to longer wavelengths by dust. However, the
hypothesis is becoming evermore constrained as the {\em FIRAS}\/ limits improve
\cite{wright93}. There is also the problem that there is no direct evidence
that VMOs can form at all, at least at the present epoch. One might still
invoke the black hole remnants of more conventional massive stars but even
these would produce a lot of background radiation \cite{bond86}.

In view of these problems, the low mass star proposal currently seems more
plausible. One reason for this is that there may be evidence that low mass
objects (LMOs) can form prolifically even at the present epoch in cooling
flows. X-ray observations of the cores of many clusters reveal the presence of
hot gas which is flowing inwards because the cooling time is less then the
Hubble time \cite{fabian84}. The mass flow rate ranges from a few $M_{\odot}$
yr$^{-1}$ to $10^{3} \; M_{\odot}$ yr$^{-1}$ and the mass appears to be
deposited over a wide range of radii with a roughly $M\propto R$ distribution.
However, the gas cannot be forming stars with the same mass spectrum as those
in the Solar neighbourhood because the central region would be brighter and
bluer than is observed \cite{fabian91}. This suggests that cooling flows
produce LMOs, possibly due to the high pressure reducing the Jeans mass. Recent
observations of large amounts of cold X-ray absorbing material in many clusters
\cite{white91} may reduce the number of LMOs required. However, one would
expect some cold gas anyway, so the implications of this are not clear.

Although cluster cooling flows could not produce the halo dark matter for
galaxies which reside outside cluster cores, Ashman \& Carr
\shortcite{ashman88} have argued that analogous high-pressure cooling flows
could have occurred on much smaller scales at an earlier cosmological epoch.
This is because, in many cosmological scenarios, one would expect the first
bound objects to be much smaller than galaxies. For example, in the
hierarchical clustering scenario, the first objects have a mass around
$10^{6} \; M_{\odot}$ and bind at a redshift in the range $20-100$; these
clouds
then cluster gravitationally to make galaxies \cite{peebles68}. A currently
popular version of this is the ``Cold Dark Matter'' model. In this case, a
detailed analysis \cite{ashman91} suggests that too much of the universe would
cool to be consistent with observation, although the fraction of gas cooling
quasi-statically (required for a cooling flow because one needs to preserve the
high pressure) might still be small. One way around this is to assume that the
first stars are very massive since their reheating effect could then limit the
fraction of the gas undergoing star formation to rather small values
\cite{thomas90}. Another  possibility is that LMOs might form prolifically even
in clouds for which the cooling time is initially much less than the dynamical
time; this could happen as a result of a ``two-phase'' instability
\cite{ashman90} whereby the amount of gas which has cooled is always such as to
ensure that the cooling time for the surviving gas is comparable to the
dynamical time. This could be relevant at either a protogalactic \cite{fall85}
or pregalactic \cite{ashman89} epoch.

The cooling flow scenarios do not specify the nature of the LMOs very precisely
but this is crucial when one comes to consider their detectability. If the dark
matter comprised ordinary hydrogen-burning stars, then they would need masses
of the order of $0.1 \; M_{\odot}$ in order to have mass-to-light ratios high
enough to explain any of the dark matter problems. However, such {\em M-dwarfs}
would still be detectable as infrared sources and present observations indicate
that their local number density cannot exceed 0.01 pc$^{-3}$ \cite{gilmore83},
which is 100 times too small to provide the local disk dark matter and 10 times
too small to provide the halo dark matter. Observations of other galaxies
support this conclusion. For example, the K-band mass-to-light ratio must
exceed 50 for NGC 456 \cite{boughn81}, 100 for M87 \cite{boughn83}, and 64 for
NGC 5907 \cite{skrutskie85}. Since the mass-to-light ratio is less than 60 for
stars bigger than $0.08 \; M_{\odot}$, the lower limit for hydrogen-burning
\cite{dantona85}, this suggests that {\em any} hydrogen-burning stars are
excluded.

Stars below $0.08 \; M_{\odot}$ are termed {\em brown dwarfs} (hereafter BDs)
and would still generate some luminosity even without nuclear burning. They
radiate first by gravitational contraction (for around $10^{7}$ years) and then
by degenerate cooling. At the present epoch, an early generation of BDs should
have extremely low luminosities, making them ideal dark matter candidates. In
determining the observational characteristics of BDs, a crucial factor is their
mass. The upper mass limit for BDs is dictated by the minimum mass required for
hydrogen-burning but the lower mass limit is much less certain. The lower mass
limit for deuterium-burning BDs is $0.015 \; M_{\odot}$ and Shu, Adams \&
Lizano \shortcite{shu87} suggest that any BDs would need to be bigger than this
since deuterium-burning may be necessary in order to generate a wind, thereby
inhibiting further accretion. However, it is not definite that BDs could
not form below $0.015 \; M_{\odot}$ and the issue of deuterium burning would be
unimportant as regards their detectability. We note that the Jeans criterion
would allow the formation of objects as small as $\sim 0.007 \; M_{\odot}$
\cite{low76} and it is possible that conditions in pregalactic or protogalactic
clouds - such as the enhanced formation of molecular hydrogen \cite{palla83} -
could allow the formation of even smaller objects. In this paper we will
consider BDs in the mass range $0.001 - 0.08 \; M_{\odot}$, the lower limit
corresponding to the transition between degenerate objects which are held
together by gravity and solid objects which are held together by electrostatic
intermolecular forces.

Objects below $0.001 \; M_{\odot}$ are sometimes termed {\em snowballs} since
they are condensations of cold hydrogen. An important lower limit on their mass
is associated with the fact that sufficiently small ones would be evaporated by
the microwave background radiation within the age of the universe. Hegyi \&
Olive \shortcite{hegyi86} first estimated the mass limit for this to be as low
as $10^{4}$ g but Phinney \shortcite{phinney85} has pointed out an error in
this
estimate and claims that the correct limit is the much higher value of
$10^{22}$ g ($\sim 10^{-11} \; M_{\odot} $). A comparable lower limit also
comes
from the number of impact parameters on the Moon and the frequency of
interstellar comets \cite{hills86}. De R\`{u}jula, Jetzer \& Mass\'{o}
\shortcite{derujula92} have shown that there may be an even more stringent
lower limit of $10^{-7} \; M_{\odot}$ since objects with masses less than this
would evaporate under their own thermal energy in a timescale less than the age
of the universe. In any case, it seems very unlikely that any fragmentation
scenario could lead to objects as small as this. Another argument against
snowballs is that, since one would only expect hydrogen to condense, less than
1/4 of the mass of the universe could go in to such objects if the primordial
helium abundance exceeds, say, 20\% \cite{phinney85}. Otherwise the current
helium abundance would be too high. We therefore neglect the possibility
of snowballs.

If the halo dark matter really does consist of LMOs, which we have seen
probably means BDs, it should be stressed that these cannot merely be the low
mass tail of the Population II mass function. Richer \& Fahlman
\shortcite{richer92} claim that the initial mass function (IMF) for
stars  in the Galactic spheroid increases steeply with decreasing mass, with no
evidence for a turnover down to $0.14 \; M_{\odot}$, and they argue that the
same could be true for halo Population II stars. However, they also point out
that the measured rotation curve of the Galaxy at the Sun requires that the
total spheroid mass not exceed $7 \times 10^{9} \; M_{\odot}$ and this implies
that the IMF cannot extend below $0.05 \; M_{\odot}$. This suggests that
Population~II LMOs fall an order of magnitude short of providing the halo dark
matter. However, there is no reason why the halo IMF should have similar
characteristics to that of the spheroid, since one would expect halo LMOs to
form at a different time and place. Indeed, if the dark baryonic matter whose
existence is suggested by cosmological nucleosynthesis constraints is locked up
in halo stars, we may {\em require} the halo IMF to extend into the LMO regime.

\subsection{Radiation from brown dwarfs}

Since it is possible that BDs provide the solution for a number of dark matter
problems, it is important to ask whether they can be detected via their
individual or collective infrared radiation. Current constraints on BDs are
rather weak (Low 1986; van der Kruit 1987; Beichman et al. 1990; Daly \&
McLaughlin 1992; Nelson, Rappaport \& Joss 1993) but the prospects of a
detection will be much better with impending space satellites such as \iso or
{\em SIRTF}\/. This problem has been addressed in various contexts by several
authors. Karimabadi \& Blitz \shortcite{karimabadi84} have calculated the
expected intensity from BDs with a discrete IMF comprising an $\Omega = 1$
cosmological background. Adams \& Walker \shortcite{adams90} have discussed the
possibility of detecting the collective emission of BDs in our own Galactic
halo for both a discrete and power law BD IMF. Daly \& McLaughlin
\shortcite{daly92} [hereafter DM] have considered the prospects of detecting
the emission of individual halo BDs of a given mass and age in the Solar
vicinity, as well as the collective emission of BDs in other galaxy halos or in
clusters of galaxies. The problem in the latter case is that the clusters have
to be at a large distance in order to fit within the satellite detector beam
and the emission would then only be detectable if the BDs were forming very
recently (viz. $z=1-2$), perhaps via cooling flows.

The work presented in this paper is complimentary to the previous studies in
several respects. Firstly, we shall be looking at the effect of an extended
mass spectrum on the BD detectability. There are no real clues as to the IMF of
BDs and we have stressed that it cannot be inferred from the IMF of younger,
more massive stars, since the formation processes responsible for Population
III BDs may have been very different. In this paper we will consider a variety
of power law IMFs in order to see what constraints can be placed from present
and future satellite observations. DM argue that assuming an extended IMF makes
it hard to assess the individual contribution of BDs of a given mass and age
for a given wavelength of observation. However, we find that this conclusion is
too pessimistic. In particular, observations at different wavelengths could
place constraints on the mass function as a whole, thereby elucidating the
relation between the BD mass function to that of more massive stars.

Another important consideration addressed in this paper the possibility that
BDs may be assembled into gravitationally bound clusters (c.f. globular
clusters) rather than being distributed uniformly throughout the halo. This is
a natural consequence of the cooling flow scenario, since one expects the
formation of {\em dark clusters} in the mass range $10^{3} - 10^{6} \;
M_{\odot}$. It is interesting that the presence of $10^{6} \; M_{\odot}$ dark
clusters in the Galactic halo could explain the observed Galactic disk heating
\cite{carr87}. Further evidence for dark clusters may come from the
line-to-continuum variations of the lensed quasar MG 2016+112
\cite{subramanian87}; the variations indicate the presence of objects in the
halo of the lensing galaxy with mass in the range $3\times 10^{4} \; M_{\odot}$
to $3\times 10^{7} \; M_{\odot}$. In any case, dynamical considerations confine
the mass and radius of such clusters to rather narrow ranges.

The plan of the paper is as follows. In \S 2 we calculate the fluxes of
the nearest individual disk and halo BDs both for a discrete IMF and an
extended IMF. In \S 3 we study the possibility that BDs are clumped into dark
clusters and consider whether the BD emission in this scenario might be easier
to detect. We also relate these results to previous work on Galactic emission
by Adams \& Walker (1990), showing how the cluster scenario may introduce
observable anisotropies in the detected emission. In \S 4 we consider the
cosmological background radiation generated by BDs, be they intergalactic (a
component perhaps of the background dark matter) or confined to galaxy halos.
Finally in \S 5 we look at the constraints which \iras already places on the BD
scenario and at the prospects that future satellites such as \iso and \sirtf
will be able to confirm it.

\section{Emission from the nearest unclustered Brown Dwarfs}

Evidence suggests that dark matter in the Solar neighbourhood may consist of
two
components: (1) halo dark matter with a local density of $\sim 0.01
 \; M_{\odot}$ pc$^{-3}$, about one-tenth that of visible matter; (2) disk dark
matter with a density of around $\sim 0.1 \; M_{\odot} $ pc$^{-3}$, comparable
to that of the visible matter. Although the evidence for halo dark matter is
strong, there is some uncertainty about the need for disk dark matter in view
of the possible existence of a thick disk \cite{kuijken89}. However, Bahcall,
Flynn \& Gould \shortcite{bahcall92} argue that a ``no disk dark matter'' model
is inconsistent with their data at the 86\% confidence level and claim a
best-fit model with a local dark matter density of $0.15 \; M_{\odot}$
pc$^{-3}$. In any case, there are likely to be {\em some} BDs in the disk,
even if they do not provide {\em all} of the disk dark matter. We will
therefore assume that BDs could contribute to either the halo or the disk dark
matter and possibly both. In this section, we will estimate the flux from the
nearest halo BD, first for a discrete IMF in \S 2.1 and then for an extended
IMF in \S 2.2. We will then make a similar estimate for the nearest disk BD in
\S 2.3. In both cases we will assume that the BDs are unclustered.

\subsection{Halo brown dwarfs with a discrete IMF}

If the halo BDs all have the same mass $m$, then the expected distance to the
nearest one depends only on the halo density $\rho_{0} = 0.01 \; M_{\odot}$
pc$^{-3}$ \cite{bahcall80} and the fraction $f_{h}$ of the halo
composed of BDs. In this case, the number density of the BDs is
   \be
      n_{0}=0.01\, f_{h}\left( \frac{ m }{M_{\odot}}\right)^{-1}\mbox{ pc}^{-3}
     \label{2.1}
   \ee
and a characteristic separation between them is $n_{0}^{-1/3}$. In order to
make a more precise estimate, we assume that the local distribution of BDs
is Poissonian. The expected distance to the nearest one is then
   \be
      \disdel = 4 \pi n_{0} \int_{0}^{\infty} r^{3} \exp [-4
      \pi n_{0} r^{3} / 3] \; dr \simeq 0.55 \, n_{0}^{-1/3} = 0.55 \,
      f_{h}^{-1/3} \mbd^{1/3} \mbox{ pc} \label{2.2}
   \ee
where $\exp [-4 \pi n_{0} r^{3} / 3]$ is the probability that a randomly placed
sphere of radius $r$ does not contain a BD and $\mbd \equiv m / 0.01 \;
M_{\odot}$. The probability that the nearest BD lies within a distance $d$
is
   \begin{eqnarray}
      P(d) & = & 4 \pi n_{0} \int_{ 0 }^{ d } r^{2} \exp [- 4 \pi n_{0} r^{3}
      / 3] \; dr =  1 - \exp [-0.71 (d/\disdel)^{3}] \nonumber \\
           & \approx & 0.7 (d/ \disdel) ^{3} \;\;\;\;\;\;\; \mbox{for } d \ll
      \disdel
      \label{2.3}
   \end{eqnarray}
and we can use this result to calculate confidence limits. For example,
satellites will have to be sensitive to BDs out to a distance $1.61 \, \disdel$
before a detection can be expected with 95\% probability.

We now determine the expected flux from the nearest BD. We will assume that BDs
in the mass range $0.001 \; M_{\odot}$ to $0.08 \; M_{\odot}$ are blackbody
radiators, which is probably a fair assumption for zero metallicity
Population~III stars, although later populations, such as those in the disk,
may be affected by the formation of solid grains on the stellar surface due to
low surface temperatures (Nelson, Rappaport \& Joss 1986; Nelson 1990). In the
absence of these effects, the observed flux at frequency $\nu$ from the nearest
BD is
   \be
      F_{\nu}^{bd}=\frac{L_{\nu}(m)}{4\pi d^{2}} = \frac{L(m)B_{\nu}(T)}
      {4 d^{2} \sigma T(m)^{4}}\label{2.4}
   \ee
where $d$ is the distance of the BD, $L_{\nu}(m)$ is the specific
luminosity and $L(m)$ is the total luminosity for a BD of mass $m$.
$B_{\nu}(T)$ is the blackbody Planck function
   \be
      B_{\nu}(T)=\frac{2h\nu^{3}}{c^{2}}\left[\exp\left(
      \frac{h\nu}{kT}\right)-1\right]^{-1} \label{2.5}
   \ee
which is dependent on the frequency  and the effective temperature $T(m)$ of
the BD. Both $T(m)$ and the $L(m)$ are dependent upon the age and opacity of
the BD. They have been calculated for a range of ages and opacities by many
authors (e.g. D'Antona \& Mazzitelli 1985; Lunine, Hubbard \& Marley 1986;
Stevenson 1986; Burrows, Hubbard \& Lunine 1989; Nelson, Rappaport \& Joss
1993). In this paper we will use the scaling relations of Stevenson
\shortcite{stevenson86}, as do DM
   \begin{eqnarray}
      L & \approx & 3.5\times 10^{-8}\mbd^{2.5}\tbd^{-1.25}\k^{0.3}\;L_{\odot}
      \nonumber \\
      T & \approx & 221\,\mbd^{0.79}\tbd^{-5/16}\k^{0.075}\mbox{ K}
      \label{2.6}
   \end{eqnarray}
Here $\tbd$ is the age of the BDs in units of $10^{10}$ years (the symbol
$\tau$ will always indicate these units) and $\k$ is the opacity in units of
$0.01$ cm$^{2}$g$^{-1}$. These relations are only valid for
   \be
      0.1 \ls \mbd \ls 10 \; \k^{-0.12} \tbd^{0.5} \label{2.7}
   \ee
since, for larger masses, non-degeneracy becomes an important factor and, for
smaller masses, one has a solid held together by atomic rather than
gravitational forces. The greatest uncertainty in these relations lies in the
value of the opacity and, in particular, in the importance of grain opacity for
BDs. The opacity probably lies in the range $1 \ls \k \ls 100$ as one goes from
Population~III to Population~I BDs but it also varies with pressure and
temperature \cite{stevenson86}. Hence the uncertainty in the opacity leads to a
factor of 4 uncertainty in luminosity and a factor of 1.5 in effective
temperature. Although in any formulae we will show the opacity dependence
explicitly, for any figures we will assume the lower limit of $\k = 1$. For the
purpose of detection, this represents the most pessimistic case and it
therefore provides the most conservative constraints.

{}From eqns (\ref{2.4}) to (\ref{2.6}) with $d$ put equal to $\disdel$, the
expected flux from the nearest BD is
   \be
      F_{\nu}^{bd} \approx 0.13 \, f_{h}^{2/3} \mbd^{1/3} f_{\nu}
      (m,\tau,\kappa) \mbox{ Jy} \label{2.8}
   \ee
where
   \be
      f_{\nu}(m,\tau,\kappa) \equiv \left[ \frac{ \fre^{3}\mbd^{-1.66} }{ \exp(
      2.17 \mbd^{-0.79} \fre \tbd^{0.31} \k^{-0.075} )-1 } \right] \label{2.9}
   \ee
with $\fre \equiv \nu/10^{13}$ Hz. The flux peaks at
   \be
      \fre\mbox{(max)} \approx 1.3\,\mbd^{0.79}\tbd^{-0.31}\k^{0.075}
      \label{2.10}
   \ee
corresponding to a wavelength
   \be
      \lambda(\mbox{max}) \approx 23\,\mbd^{-0.79}\tbd^{0.31}\k^{-0.075}
      \mbox{ }\mu\mbox{m} \label{2.11}
   \ee
with a value
   \be
      F_{\nu}^{bd} \mbox{(max)} \approx 18 \, f_{h}^{0.67} \mbd^{1.04} \tbd^{
      -0.94} \k^{0.225} \mbox{ mJy} \label{2.12}
   \ee
Figure (1) shows $F_{\nu}^{bd}$ as a function of $\nu$ for values of $m$ from
$0.001 \; M_{\odot}$ to $0.08 \; M_{\odot}$. It assumes that $f_{h} = 1$ and
that the BDs all have the same age of 10 Gyr. We stress that the curves only
indicate the {\em most likely} fluxes since, for given $m$ and $f_{h}$, there
will be a distribution in $F_{\nu}^{bd}$ associated with the distribution in
$d$
given by eqn (\ref{2.3}). For this reason, unless one has an independent
measure of the BD temperature, a positive detection would not uniquely specify
the values of $m$ and $f_{h}$ but only a likelihood distribution of their
values.

DM also calculate the peak frequency and flux but not the whole spectrum
because they approach the problem somewhat differently. For observations at a
given frequency, they determine the mass of the BD whose flux peaks there
[given by eqn (\ref{2.10})] and, for a given detector sensitivity and sky
coverage, they infer constraints on the fraction of the halo in BDs of this
mass. In principle, observations at a given frequency allow constraints on BDs
with masses different from that specified by eqn (\ref{2.10}): eqns (\ref{2.8})
and (\ref{2.9}) imply that the constraint on the density of BDs first decreases
as $\exp (m^{-0.79})$ and then increases as $m^{2}$ with increasing $m$.
However, as we now discuss, the situation may be more complicated than this if
the BDs span a range of masses.

\subsection{Halo brown dwarfs with an extended IMF}

The assumption of a discrete IMF is clearly unrealistic since observations show
that stars generally have an extended IMF. This modifies the previous
discussion in several ways. Firstly, one cannot necessarily assume that the BD
mass which dominates the flux at a particular frequency is still given by eqn
(\ref{2.10}). While the constraints derived above are still valid, they may be
too conservative. Previous authors have assumed that the effects of an extended
IMF could completely mask the effects of other parameters. Indeed, this is why
DM only use the observations at a given frequency to constrain the BDs of a
particular mass and age. However, we will show that this is overly pessimistic
because eqn (\ref{2.8}) may still be approximately valid in some circumstances
even in the extended case. The second modification is a positive one in that
observations at {\em different} frequencies would in principle convey {\em
information} about the IMF, so it is important to consider how this could be
extracted.

For an extended IMF, one can only define the distance to the nearest BDs over a
non-zero mass range since the number of BDs of any particular mass is formally
zero. In order to deal with this problem, we divide the mass range into $N$
logarithmically equal strips and approximate the continuous IMF by a series of
step functions. If $m_{l}$ and $m_{u}$ are the lower and upper mass cut-offs,
respectively, the $n$th strip is then $\{ \alpha^{n} m_{l} < m <
\alpha^{n+1} m_{l} \}$ where $n$ goes from 0 to $N-1$ and $m_{u} = m_{l}
\alpha^{N}$. We assume $N$ is sufficiently large that $m$ can be regarded as
constant within each strip, so that the IMF is essentially represented by a sum
of delta functions. We then calculate the flux from the nearest BD in each
strip independently, and, for each frequency band, find the strip which
dominates. The validity of this approximation depends crucially on the value of
$\alpha$: it must be large enough that the flux in each frequency band can be
assumed to be dominated by a single strip but not so large that the assumption
of constant $m$ within each strip gives a poor estimate of the flux there. We
now consider the condition for this in more detail.

If the IMF is denoted by $\phi (m)$ so that the number density of BDs in the
mass interval $(m,m + dm)$ is $dn = \phi(m) \, dm$, then the fraction of the
total BD density comprising BDs in the mass interval $(m , \alpha m)$ is
   \be
      g(m) \equiv \frac{\ts \int_{m}^{\alpha m} m \phi (m) \; dm }{\ts
      \int_{m_{l}}^{m_{u}} m \phi (m) \; dm } \label{2.13}
   \ee
where $\phi$ is normalised to an assumed local BD halo density of
   \be
      \rho_{0} = \int_{m_{l}}^{m_{u}} m \phi(m) \; dm = 0.01 \, f_{h} \;
      M_{\odot} \mbox{ pc}^{-3} \label{2.14}
   \ee
The local number density of BDs in the mass interval $(m,\alpha m)$ is then
   \be
      n(m) = \int_{m}^{\alpha m} \phi (m) \; dm \simeq \frac{ g(m) \rho_{0} }{
      m } \label{2.15}
   \ee
where all the BDs are taken to have the mass $m$. The expected distance to the
nearest BD in the mass interval $(m, \alpha m)$ is then
   \be
      \dis \simeq 0.55 \, f_{h}^{-1/3} g(m)^{-1/3} \mbd^{1/3} \mbox{ pc}
      \label{2.17}
   \ee
and the associated flux is
   \be
      F_{\nu}^{bd} \approx 0.13 \, f_{h}^{2/3} g(m)^{2/3} \mbd^{1/3} f_{\nu}
      (m,\tau,\kappa) \mbox{ Jy} \label{2.18}
   \ee
These equations reduce to eqns (\ref{2.2}) and (\ref{2.8}) if one puts $g(m) =
1$. In principle, given the form of $g(m)$, one can now find the value of $m$
which maximises the flux at any particular frequency but one must ensure that
$\alpha$ is large enough for the strip $(m, \alpha m)$ to provide a good
estimate of that flux. Since the
errors in the flux associated with the uncertainty in $\k$, the deviation of
the actual value of $d$ from $\dis$ and the inaccuracies in the scaling laws of
eqns (\ref{2.6}) are around a factor of 2, any value of $\alpha$ which produces
errors smaller than this will suffice.

In order to determine a suitable value for $\alpha$, we need to specify the
IMF. A simple and widely used IMF is the power law form:
   \be
      \phi(m) = K \left( \frac{ m_{l} }{ m } \right)^{x} \;\;\;\;\;\;\;\;
      (m_{l} \leq m \leq m_{u}) \label{2.19}
   \ee
where eqn (\ref{2.2}) requires
   \be
      K=\left\{\begin{array}{ll}
        0.01\, f_{h} \left( \frac{\ts 2-x}{\ts \beta^{2-x}-1} \right) \left(
        \frac{\ts m_{l}}{\ts M_{\odot}} \right)^{-2} \; M_{\odot}^{-1}
        \mbox{ pc}^{-3} & (x\neq 2) \\
        0.01\, f_{h} \left( \frac{\ts 1}{\ts \ln \beta} \right) \left(
\frac{\ts
        m_{l}}{\ts M_{\odot}} \right)^{-2} \; M_{\odot}^{-1}
        \mbox{ pc}^{-3} & (x=2)
      \end{array}\right. \label{2.20}
   \ee
with $\beta \equiv m_{u} / m_{l}$. The function $g(m)$ scales as $m^{2-x}$, so
most of the density is in the smallest BDs for $x > 2$ and in the largest ones
for $x < 2$. There are no theoretical grounds for believing that the stellar
IMF should be described by a power law but observations suggest that it is a
good approximation at least over limited mass ranges. Although one could
consider more general forms for the IMF, there is little purpose in this since
observations at a limited number of frequencies could never determine the extra
parameters involved, so one might as well assume the simplest.

What values of $x$ are reasonable? Luminous matter in the Solar neighbourhood
(both Population~I and Population~II) can be described by a power law IMF with
$x \sim 2.7$ for $2 \; M_{\odot} \ls m \ls 10 \; M_{\odot}$ \cite{scalo86}.
Above $10 \; M_{\odot}$ the IMF is very uncertain and below $2 \; M_{\odot}$ it
appears to be bimodal with peaks at $0.3 \; M_{\odot}$ and $1.2 \; M_{\odot}$.
Richer \& Fahlman \shortcite{richer92} claim that stars in the Galactic
spheroid have $x=4.5\pm 1.2$ below $0.5 \; M_{\odot}$ down to at least $0.14 \;
M_{\odot}$. The value of $x$ below $0.1 \; M_{\odot}$ (which is relevant to
BDs) is very uncertain since such stars are hard to observe. Even if it
were known for stars forming at the present epoch, there is no reason to
suppose that the same value would apply for stars forming at earlier times. The
value of $x$ relevant to halo BDs is therefore highly uncertain and we
will consider values in the range 0 to 4.

Eqns (\ref{2.18}) and (\ref{2.19}) imply that, for a given $\nu$, the
contribution to the flux from BDs of different masses first increases as
$\exp(-m^{-0.79})$ and then goes as $m^{0.79-2x/3}$ with increasing $m$. This
shows that the {\em highest} mass $m_{u}$ dominates at all $\nu$ for $x < 1.2$.
In the more likely situation with $x > 1.2$, the dominant mass is still roughly
given by eqn (\ref{2.10}), although it is multiplied by a factor which depends
on $x$. Thus the notion exploited by DM in the discrete IMF case, that each
frequency is dominated by a particular mass, still applies. If one is
approximating the IMF by a series of step functions, as we are, this means that
each mass strip corresponds to a particular frequency band unless $x < 1.2$.

We examine how to choose the value of $\alpha$ for a power law IMF in the
Appendix. For our purposes, it is appropriate to take the number of strips to
be $N = 10$. The associated values of $\alpha$ are then 1.55 for $m_{l} = 0.001
\; M_{\odot}$ and $m_{u} = 0.08 \; M_{\odot}$ (case a), 1.23 for $m_{l} = 0.01
\; M_{\odot}$ and $m_{u} = 0.08 \; M_{\odot}$ (case b), and 1.26 for $m_{l} =
0.001 \; M_{\odot}$ and $m_{u} = 0.01 \; M_{\odot}$ (case c). There is no
reason, in principle, why the mass cut-offs in $\phi$ should correspond to the
theoretical BD mass range of $0.001 - 0.08 \; M_{\odot}$, so these values for
$m_{l}$ and $m_{u}$ are chosen to represent three distinct situations: in (a)
the IMF is as wide as possible; in (b) it spans only the highest possible mass
range; in (c) it spans only the lowest possible mass range.

Figures (2a) to (2c) indicate the likely flux at different frequencies for
various values of $x$ in these three cases. The discrete IMF cases for $m =
m_{u}$ (curve A) and $m = m_{l}$ (curve B) are also included for comparison.
The curves are derived by superposing the fluxes corresponding to the different
mass strips and taking the envelope. They should not be interpreted as spectra
since an {\em individual} BD always has a blackbody spectrum. They merely
indicate the sensitivity required at each frequency in order to detect the
nearest BD which dominates at that frequency. For large values of $x$, the
contribution of different mass ranges can be identified by the kinks. These
kinks are not physical but indicate the errors involved in approximating the
IMF by step functions. The kinks become less noticeable as $x$ decreases since
the errors are reduced. They disappear altogether for $x < 1.2$ since, as
already discussed above, the maximum mass $m_{u}$ then dominates at {\em every}
frequency. Increasing $m_{l}$ or decreasing $m_{u}$ reduces the frequency range
of the backgrounds corresponding to cases (A) and (B).

\subsection{Disk brown dwarfs}

The calculations of the previous section can also be applied to disk BDs,
although these are likely to be younger $(\sim 5$ Gyr) and have higher
opacities $(\sim 1$ cm$^{2}$ g$^{-1}$) than their halo counterparts. We will
assume a disk dark matter density of $0.15 \; M_{\odot}$ pc$^{-3}$
\cite{bahcall92}, so that
   \be
      \rho_{0} = \int_{m_{l}}^{m_{u}} m \phi(m) \; dm = 0.15 \, f_{d} \;
      M_{\odot} \mbox{ pc}^{-3} \label{2.21}
   \ee
where $f_{d}$ is the fraction of the disk dark matter in BDs. Eqns (\ref{2.2})
and (\ref{2.15}) then give the expected distance to the nearest disk BD in the
mass range $(m, \alpha m)$ as
   \be
      \dis \simeq 0.22 \, f_{d}^{-1/3} g(m)^{-1/3} \mbd^{1/3} \mbox{ pc}
      \label{2.22}
   \ee
where the distribution around $\dis$ is given by eqn (\ref{2.3}). Using eqns
(\ref{2.4}) to (\ref{2.6}), we obtain an expected flux from the nearest disk BD
in the range $(m, \alpha m)$ of
   \be
      F_{\nu}^{bd} \approx 0.8 \, f_{d}^{2/3} g(m)^{2/3} \mbd^{1/3} f_{\nu}
      (m,\tau,\kappa) \mbox{ Jy} \label{2.23}
   \ee
For a discrete IMF, $g(m) = 1$ and the expected peak flux is
   \be
      F_{\nu}^{bd} \mbox{(max)} \approx 0.1 \, f_{d}^{0.67} \mbd^{1.04} \tbd^{
      -0.94} \k^{0.225} \mbox{ Jy} \label{2.24}
   \ee
Figure (3a) shows the spectra in this case. For a power law IMF with $m_{l} =
0.001 \; M_{\odot}$ and $m_{u} = 0.08 \; M_{\odot}$, the envelopes are shown by
Figure (3b) where the curves are to be interpreted as before. The figures are
analogous to Figures (1) and (2a) in the previous section except we have taken
$\tbd = 0.5$ and $\k = 100$, appropriate for disk population stars. As one
would expect, disk BDs are considerably brighter than halo BDs due to their
higher density and greater intrinsic luminosity.

Nelson, Rappaport \& Joss (1993) have also considered the emission from disk
BDs with a power law IMF. However, they assume that the BDs are forming over a
range of epochs with a prescribed formation rate, which is clearly better than
our approximation that the formation is coeval. They assume that $m_{u} = 0.085
\; M_{\odot}$ but allow different values of $m_{l}$ and $x$. It should be
stressed that Nelson et al. do not constrain the BDs at $m_{u}$ to have the
observed density, so they are not treating the BD population as an extension of
the IMF of the visible stars. It may therefore be inappropriate to take $m_{u}$
as high as $0.085 \; M_{\odot}$.

If all of the disk stars
formed under similar physical conditions, one {\em would} expect the BD IMF to
be a continuation of the IMF of the more massive stars.
In this context, it should be stressed that there are likely to be {\em some}
BDs in the disk, since we see stars down to $0.08 \; M_{\odot}$ and there is no
reason why the minimum stellar fragmentation mass should be the same as the
minimum mass for hydrogen-burning stars. Indeed, if the IMF of visible stars in
the Solar neighbourhood can be extrapolated into the BD mass range, then they
necessarily dominate the number density for $x > 1$. Tinney, Mould \& Reid
\shortcite{tinney92} find that disk stars close to the hydrogen-burning limit
have a number density per unit mass below $0.2 \; M_{\odot}^{-1}$pc$^{-3}$.
Eqn (\ref{2.19}) then gives
   \be
      \phi(m_{u}) = K \beta^{-x} \ls 0.2 \; M_{\odot}^{-1}
      \mbox{pc}^{-3} \label{2.25}
   \ee
where $\beta \equiv m_{u}/m_{l}$ and eqn (\ref{2.21}) implies
   \be
      K=\left\{\begin{array}{ll}
        0.15 \,f_{d} \left( \frac{\ts 2-x}{\ts \beta^{2-x}-1} \right) \left(
        \frac{\ts m_{l}}{\ts M_{\odot}} \right)^{-2} \; M_{\odot}^{-1}
        \mbox{pc}^{-3} & (x \neq 2) \\
        0.15 \, f_{d} \left( \frac{\ts 1}{\ts \ln \beta} \right) \left(
        \frac{\ts m_{l}}{\ts M_{\odot}} \right)^{-2} \; M_{\odot}^{-1}
        \mbox{pc}^{-3} & (x = 2)
      \end{array}\right. \label{2.26}
   \ee
Note that $m_{l}$ and $m_{u}$ need not be the same as in the halo case and that
the value of $x$ is treated as unknown. On the basis of the observed IMF, one
would therefore expect the BD contribution to the dark matter to satisfy
   \be
      f_{d} \ls 8.5 \times 10^{-3} \; \left( \frac{\ts 1 - \beta^{x-2}}{\ts
2-x}
      \right) \left( \frac{m_{u}}{0.08 \; M_{\odot}} \right)^{2} \;\;\;\;\;\;\;
      \;\;\;\;\;\;\;\;\; (x \neq 2) \label{2.27}
   \ee
In this case, it is reasonable to assume $m_{u} = 0.08 \; M_{\odot}$ since the
BDs are just the tail of the standard IMF. If disk BDs have a Salpeter mass
function $(x = 2.3)$, then, taking $m_{u} = 0.08 \; M_{\odot}$ and $m_{l} =
0.001 \; M_{\odot}$, we find that disk BDs could make up at most 8\% of the
disk dark matter. If we want $f_{d} = 1$, then we must either increase $x$ to
3.1 or decrease $m_{l}$ to $5 \times 10^{-7} \; M_{\odot}$. The second option
seems implausible since this value for $m_{l}$ is well below calculated values
for the minimum Jeans mass, although it is intriguingly close to the
De~R\`{u}jula et al. (1992) estimate of the minimum possible mass.

\section{Emission from halo Brown Dwarf clusters}

As discussed in \S 1, it is possible that halo BDs may be clumped into dark
clusters (c.f. globular clusters). In this section we investigate the
detectability of such clusters and look at the implications that it may have
for the detection of a general Galactic background \cite{adams90}. We also
compare the present and possible future constraints imposed from infrared
satellites such as \iras and \iso to the dynamical constraints on BD clusters.

\subsection{Dynamical constraints}

We assume that they all have the same mass $M_{c}$ and radius $R_{c}$. An upper
limit on $M_{c}$ derives from the requirement that the halo objects must not
heat up the disk stars too much:
   \be
      M_{c} < 3 \times 10^{6} \tau_{g}^{-1} \; M_{\odot} \label{3.1}
   \ee
where $\tau_{g} \equiv t_{g} / 10$ Gyr is the age of the Galaxy \cite{lacey85}.
If the halo objects were single black holes, they would need to have
   \be
      M_{c} < 7 \times 10^{4} \tau_{g}^{-1} \left( \frac{ a }{ \mbox{2 kpc}
      } \right) \; M_{\odot} \label{3.2}
   \ee
where $a$ is the halo core radius, in order to avoid too much mass accumulating
in the Galactic centre through dynamical friction \cite{carr93}. However, this
need not apply if they were BD clusters because such clusters would be
disrupted by collisions before dynamical friction could operate if they had a
radius satisfying \cite{carr87}
   \be
      R_{c} > 1.4 \; \tau_{g}^{-1} \left( \frac{ a }{ \mbox{2 kpc} }
\right)^{2}
      \mbox{ pc} \label{3.3}
   \ee
In order to avoid evaporating by now as a result of 2-body relaxation, one
would also require the clusters to have
   \be
      R_{c} \gs 0.06 \; \mbd^{2/3} \tau_{g}^{2/3} \mc^{-2/3} \mbox{ pc}
      \label{3.4}
   \ee
where $\mc \equiv M_{c} / 10^{6} \; M_{\odot}$. An upper limit on $R_{c}$ comes
from requiring that the clusters avoid disruption at our own Galactocentric
radius $R_{0} = 8.5$ kpc, so they can be observed locally, and this implies
   \be
      R_{c} < 25 \; \tau_{g}^{-1} \left( \frac{ R_{0} }
      { \mbox{8.5 kpc} } \right)^{2} \mbox{ pc} \label{3.5}
   \ee
These dynamical limits are indicated by the bold lines in Figure (4), which
shows that the values of $M_{c}$ and $R_{c}$ are constrained to a rather narrow
range. If one merely requires that the clusters do not disrupt at the {\em
edge} of the halo, the upper limit of eqn (\ref{3.5}) is increased by a factor
of $(R_{h} / R_{0})^{2}$ where $R_{h}$ is the halo radius. For $R_{h} = 100$
kpc the dynamically permitted region is indicated by the dotted line in Figure
(4). The dynamical friction limits are sensitive to the value of $a$: we show
the limits for $a = 2$ kpc and $a = 8$ kpc since this spans the range of likely
values. The evaporation limit given by eqn (\ref{3.4}) depends on the value of
$m$: Figure (4) corresponds to $0.02 \; M_{\odot}$, which turns out to optimize
the prospects of detection.

Disk dark matter could not comprise clusters of significant mass since objects
larger than $2 \; M_{\odot}$ would disrupt binary systems in the disk
\cite{bahcall85}. Although the figure of $2 \; M_{\odot}$ has been disputed
\cite{weinberg88}, the correct figure is unlikely to exceed about $10 \;
M_{\odot}$. BDs in a cosmological background could, in principle, be in
clusters much larger than $10^{6} \; M_{\odot}$. Such objects would, most
naturally, be interpreted as dark dwarf galaxies and it is not inconceivable
that most of the dark baryons could be contained in them. We discuss this
possibility further in \S 5.

\subsection{The flux from individual clusters}

If the halo BDs reside in clusters, one might at first anticipate them being
easier to detect than BDs in the unclustered scenario. This is because,
although the distance to the nearest source is increased by a factor
$(M_{c}/m)^{1/3}$, the luminosity is increased by $M_{c}/m$, giving an increase
in flux of $(M_{c}/m)^{1/3}$. For example, if the halo is composed of $0.01 \;
M_{\odot}$ BDs, the nearest source will be 500 times brighter if in clusters of
$10^{6} \; M_{\odot}$ than if unclustered. However, this argument only applies
if the nearest clusters can be treated as point sources rather than extended
objects. Whether this is the case will depend upon its distance and size. Since
the local cluster number density is $n_{c}(M_{c}) = 10^{-8} f_{h} \mc^{-1}$
pc$^{-3}$, eqn (\ref{2.2}) with $m$ replaced by $M_{c}$ shows that the expected
distance to the nearest one is
   \be
      \cdis = 260 \, f_{h}^{-1/3} \mc^{1/3} \mbox{ pc} \label{3.6}
   \ee
and the probability of its being closer than this is given by eqn (\ref{2.3})
with $\disdel$ being replaced by $\cdis$. The cluster will be observable as a
point source provided its angular size is less than the angular resolution
$\theta_{res}$ of the telescope. The angular size of a cluster at the distance
given by eqn (\ref{3.6}) is
   \be
      \theta_{c} = 27 \left( \frac{ R_{c} }{ \mbox{pc} } \right) \left(
      \frac{ f_{h} }{ \mc } \right)^{1/3} \mbox{ arcmin} \label{3.7}
   \ee
Figure (4) show that the clusters would have $R_{c} > 0.1$ pc and $\mc \ls 1$,
so eqn (\ref{3.7}) implies that the nearest ones would have angular diameters
of at least 3 arcmins for $f_{h} = 1$. The resolution of the IRAS Faint Source
Survey plates at 12 $\mu$m is 15 arcsec \cite{moshir92} and the resolutions of
ISO and SIRTF are expected to be 6 arcsec and 3 arcsec respectively (Cesarsky
1992; Werner 1992), so it seems that the nearest clusters are likely to be
extended sources. This is illustrated by the dotted line in the top left hand
corner of Figure (4), which gives the division between point-like and extended
sources for $\theta_{res} = 15$ arcsec. Comparison with the dynamical limits
show that the clusters must lie well within the extended regime.

In the extended case, the intensity of a BD cluster is
   \be
      I_{\nu}^{clus}=\frac{M_{c} \Gamma_{\nu}}{4 \pi^{2} R_{c}^{2}}
      \label{3.8}
   \ee
where $\Gamma_{\nu}$ is the specific luminosity per mass. The intensity has no
dependence on distance until the cluster is so far away that it cannot be
resolved. The critical distance is
   \be
      D = 28 \left( \frac{ R_{c} }{ \mbox{pc} } \right) \left( \frac{
      \theta_{res} }{ \mbox{15 arcsec} } \right)^{-1} \mbox{ kpc} \label{3.9}
   \ee
and this is more than the halo radius, so that all halo clusters are
extended, if
   \be
      R_{c} > 3.6 \; \left( \frac{ \theta_{res} }{ \mbox{15 arcsec} } \right)
      \left( \frac{ R_{h} }{ \mbox{100 kpc} } \right) \mbox{ pc} \label{3.10}
   \ee
In order to determine the fraction of the sky in which one measures the
intensity (\ref{3.8}), we need to determine the covering factor of the extended
clusters. If $D$ is sufficiently small that the cluster density can be
approximated by its local value $n_{c}$, then the covering factor is
   \be
      K_{e} = \pi n_{c} R_{c}^{2} D  = 8.6 \times 10^{-4} \left( \frac{ f_{h} }
      { \mc } \right) \left( \frac{ R_{c} }{ \mbox{pc} } \right)^{3}
      \left( \frac{ \theta_{res} }{ \mbox{15 arcsec} } \right)^{-1}
\label{3.11}
   \ee
This formula only applies for $D < 3$ kpc (say), which from eqn (\ref{3.10})
requires $R_{c} < 0.1$ pc. This would exceed 1 for
   \be
      M_{c} < 860 \; f_{h} \left( \frac{ R_{c} }{ \mbox{pc} } \right)^{3}
      \left( \frac{ \theta_{res} }{ \mbox{15 arcsec} } \right)^{-1} \;
      M_{\odot}  \label{3.12}
   \ee
but dynamical constraints preclude this, so the extended sources never overlap
in this case. If the critical distance $D$ in eqn (\ref{3.9}) exceeds $R_{h}$,
$K_{e}$ is replaced by $K_{h}$, the covering factor of {\em all} the halo
clusters, and this is calculated in \S 3.3 [see eqn (\ref{3.24})]. For
intermediate values of $D$ (viz. 3 kpc $< D <$ 8.5 kpc), the covering factor of
the extended clusters is somewhat less than $K_{h}$. In all three cases, there
will be a general Galactic background if $K_{h} > 1$, with the intensity
everywhere exceeding the value given by eqn (\ref{3.8}) by the factor $K_{h}$
for $D > R_{h}$. Figure (4) shows that this applies in at least some of the
permitted domain. If $K_{h} < 1$, one is only observing a single cluster and
the intensity (\ref{3.8}) is only seen over a fraction of the sky. We confine
attention to this case for the rest of this subsection and consider the other
case in \S 3.3.

If the BDs comprising the cluster have a discrete mass spectrum, $\Gamma_{\nu}$
just depends on $m$ and $t$ and the expected intensity of any extended
cluster is
   \be
      I_{\nu}^{clus} \approx 1.2 \, \mc f_{\nu}(m,\tau,\kappa) \left( \frac{
      R_{c} }{ \mbox{pc} } \right)^{-2} \mbox{ MJy sr}^{-1} \label{3.13}
   \ee
where $f_{\nu}(m,\tau,\kappa)$ is defined by eqn (\ref{2.9}). Equation
(\ref{3.13}) peaks at the value of $\nu$ given by (\ref{2.10}) with a value
   \be
      I_{\nu}^{clus} \mbox{(max)} \approx 0.2 \, \mc \mbd^{0.71} \tbd^{-0.94}
      \k^{0.225} \left( \frac{ R_{c} }{ \mbox{pc} } \right)^{-2} \mbox{ MJy sr}
      ^{-1} \label{3.14}
   \ee
Figure (5a) shows the expected intensity from extended clusters of mass $M_{c}
= 10^{6} \; M_{\odot}$ and radius 1 pc for various BD masses, assuming a local
halo density of $0.01 \; M_{\odot}$pc$^{-3}$ and various values of $m$.

If the BDs have a continuous mass spectrum, as described by equation
(\ref{2.19}), then
   \be
      \Gamma_{\nu} = \frac{\ts \int_{m_{l}}^{m_{u}} L_{\nu} (m,t) \phi(m)
      \; dm}{\ts \int_{m_{l}}^{m_{u}} m \phi(m) \; dm} \label{3.15}
   \ee
[We assume that the clusters are large enough to include the entire mass range
$m_{l}$ to $m_{u}$. If the clusters  formed out of a pre-existent population of
BDs with masses between $m_{l}$ and $m_{u}$, then they would need to exceed a
critical size in order to contain BDs all the way up to $m_{u}$ (for $x>2$) or
all the way down to $m_{l}$ (for $x<2$). This requires $M_{c} > m_{u}^{x-1}
m_{l}^{2-x}$ (for $x > 2$) or $M_{c} > m_{l}^{x-1} m_{u}^{2-x}$ (for $x < 2$)
and these conditions are likely to be satisfied.] From eqns (\ref{2.4}),
(\ref{2.6}), (\ref{2.19}) and (\ref{3.8}), the intensity of any extended
cluster is
   \be
      I_{\nu}^{clus} \approx 1.2 \, \mc S_{\nu}(\tau,\kappa) \left( \frac{
      R_{c} }{ \mbox{pc} } \right)^{-2} \mbox{ MJy sr}^{-1} \label{3.16}
   \ee
where
   \be
      S_{\nu}(\tau,\kappa) = \alpha \int_{m_{l}}^{m_{u}} f_{\nu}(m,\tau,\kappa)
      m^{1-x} \; dm \label{3.17}
   \ee
with
   \be
      \alpha=\left\{\begin{array}{ll}
                \left( \frac{\ts 2-x}{\ts a^{2-x}-1} \right) \left( \frac{\ts
                m_{l}^{}}{\ts 0.01M_{\odot}} \right)^{x-2} & (x\neq 2) \\
                \left( \frac{\ts 1}{\ts \ln a} \right) & (x=2)
      \end{array}\right. \label{3.18}
   \ee
Eqn (\ref{3.16}) also covers the discrete IMF case if we replace
$S_{\nu}(x,\tau,\kappa)$ with $f_{\nu}(m, \tau, \kappa)$. Figure (5b) gives the
expected intensity for power law IMFs with different values of $x$ if $m_{u} =
0.08 \; M_{\odot}$ and $m_{l} = 0.001 \; M_{\odot}$. Curves (A) and (B)
correspond to discrete BD mass spectra with $m = 0.08 \; M_{\odot}$ and $m =
0.001 \; M_{\odot}$, respectively. As expected, for small values of $x$ the
spectra tend to (A), while for large values of $x$ they tend to (B).

\subsection{Background emission from halo clusters}

So far we have dealt with observations of individual BDs or BD clusters but we
can also detect BDs in our own halo by looking for {\em extended} infrared
emission over the whole sky. Here we calculate the expected intensity of such
emission. As mentioned in \S 2, if the local disk dark matter is also composed
of BDs, then we may require some means of discriminating between halo and disk
BDs and searching for high Galactic latitude emission provides this method.
Adams \& Walker \shortcite{adams90} and  DM have looked at the problem for
unclustered BDs but our analysis is more general in that it includes the
clustered case.

We assume that the halo is spherically symmetric and that the density at a
distance $r$ from the Galactic centre is
   \be
      \rho (r)=\left\{ \begin{array}{ll}
          \rho_{c} \left( \frac{\ts a^{2}}{\ts a^{2}+r^{2}} \right) &
          (r\leq R_{h}) \\
          0 & (r>R_{h})
      \end{array}\right. \label{3.19}
   \ee
where $a$ is the halo core radius, $\rho_{c}$ is the core density and $R_{h}$
is the cut-off radius. Since the local halo density $\rho_{0} \simeq 0.01 \;
M_{\odot}$pc$ ^{-3}$, the value of the core density is related to $a$ by
   \be
      \rho_{c} = \left[ 0.01 + 0.72 \left( \frac{ a }{ \mbox{kpc} } \right)
      ^{-2} \right] \mbox{ } M_{\odot} \mbox{pc}^{-3} \approx 0.7 \left( \frac{
      a }{ \mbox{kpc} } \right)^{-2} \mbox{ } M_{\odot} \mbox{pc}^{-3}
      \label{3.20}
   \ee
where the last formula applies if $a$ is much less than the Galactocentric
radius $R_{0} = 8.5$ kpc. If BDs only comprise a fraction $f_{h}$ of the
halo, we will assume that their density is everywhere $f_{h}$ times $\rho(r)$,
although they could in principle have a different distribution.

The intensity of a halo composed entirely of unclustered BDs has been
calculated by Adams \& Walker \shortcite{adams90} for the case in which $a = 2$
kpc, corresponding to a value for $\rho_{c}$ of $0.19 \; M_{\odot} $pc$^{-3}$
in
eqn (\ref{3.20}). However, the core radius is not well determined and could be
anywhere between 1 and 4 kpc, so we will keep it as a free parameter. The
intensity can then  be expressed as
   \be
      I_{\nu}^{halo}=\frac{\Gamma_{\nu}}{4\pi} \rho_{0} R_{0} J(b,l,a,R_{h})
      \left[ 1 + \left( \frac{ a }{ R_{0} } \right)^{2} \right] \label{3.21}
   \ee
where $\Gamma_{\nu}$ is the specific intensity defined by eqn (\ref{3.15}) and
$J(b,l,a,R_{h})$ is a dimensionless function of the Galactic coordinates $b$
and $l$, the core radius $a$, and the halo radius $R_{h}$:
   \be
      J(b,l,a,R_{h}) \equiv \frac{ 1 }{\sqrt{1+\alpha^{2}-\mu^{2}}} \left[
      \tan^{-1} \left( \frac{\beta-\mu}{\sqrt{1 + \alpha^{2} - \mu^{2}}}
\right)
      + \tan^{-1} \left( \frac{\mu}{\sqrt{1+\alpha^{2}-\mu^{2}}} \right)
      \right] \label{3.22}
   \ee
with $\alpha \equiv a/R_{0}$, $\beta \equiv R_{h}/R_{0}$ and $\mu \equiv \cos b
\cos l$. Note that the dependence on $a$ in eqn (\ref{3.21}) is weak provided
$a
\ll R_{0}$. Adams \& Walker assume a halo with no cut-off, in which case $\beta
\rightarrow \infty$ and the first term inside the square brackets in eqn
(\ref{3.22}) becomes $\pi/2$; the dependence on $R_{h}$ is anyway weak for
$R_{h} \gg R_{0}$. For a power law BD IMF, the specific luminosity per mass is
   \be
      \Gamma_{\nu} \approx 2.3 \times 10^{-18} S_{\nu}(\tau,\kappa)
      \mbox{ erg s$^{-1}$Hz$^{-1}$g}^{-1}
      \label{3.23}
   \ee
where $S_{\nu}(\tau,\kappa)$ is given by eqn (\ref{3.17}), so the observed
intensity will be
   \be
      I_{\nu}^{halo} \approx 330 \, f_{h} J(b,l,a,R_{h}) S_{\nu}
      (\tau,\kappa)\mbox{ Jy sr}^{-1} \;\;\;\;\;\;  (a \ll R_{0}) \label{3.24}
   \ee
For a discrete BD mass spectrum, $S_{\nu}(\tau, \kappa)$ is replaced by
$f_{\nu}(m, \tau, \kappa)$ and the peak flux is
   \be
      I_{\nu}^{halo}\mbox{(max)}\approx 45 \, f_{h} J(b,l,a,R_{h}) \mbd^{0.71}
      \tbd^{-0.94} \k^{0.225} \mbox{ Jy sr}^{-1} \;\;\;\;\;\;\;\; (a \ll R_{0})
      \label{3.25}
   \ee

If the BDs are clustered, the situation is more complicated. As far as the
detectability of our halo emission is concerned, it does not matter whether the
BDs are clustered provided the clusters cover the sky. Since the halo covering
factor is
   \begin{eqnarray}
      K_{h} & = & \frac{ I_{\nu}^{halo} }{ I_{\nu}^{clus} } = \pi
      R_{c}^{2} \rho_{0} R_{0} M_{c}^{-1} J(b,l,a,R_{h}) \left[ 1 + \left(
      \frac{ a }{ R_{0} } \right)^{2} \right] \nonumber \\
            & = & 2.6 \times 10^{-4} \; f_{h} J(b,l,a,R_{h}) \mc^{-1} \left(
      \frac{ R_{0} }{ \mbox{8.5 kpc} } \right) \left( \frac{ R_{c} }{
      \mbox{pc} } \right)^{2} \left[ 1 + \left( \frac{ a }{ R_{0} } \right)^{2}
      \right] \label{3.26}
   \end{eqnarray}
the condition for this is
   \be
      R_{c}\gs 60\; J(b,l,a,R_{h})^{-1/2} \left( \frac{ \mc }{ f_{h} }
      \right)^{1/2} \left( \frac{ R_{0} }{ \mbox{8.5 kpc} } \right)^{-1/2}
      \left[ 1 + \left( \frac{ a }{ R_{0} } \right)^{2} \right]^{-1/2}
      \mbox{ pc} \label{3.27}
   \ee
In this case, the intensity of the Galactic background has the same frequency
dependence as that of an individual cluster but it is enhanced by the factor
$K_{h}$. Note that eqn (\ref{3.26}) assumes that all the halo clusters are
extended, so that $I_{\nu}^{clus}$ is given by eqn (\ref{3.8}). Otherwise eqn
(\ref{3.26}) is an underestimate since it only gives the covering factor of the
extended clusters $K_{e}$. If we assume $a = 2$ kpc, $R_{0} = 8.5$ kpc and
$R_{h} = 100$ kpc, then the cluster radius would have to be at least $60 \;
\mc^{1/2}$ pc for clusters to cover the sky in the direction of the Galactic
anticentre ($b=0$, $l=\pi$). In this case, they would cover it in {\em every}
direction. The line corresponding to eqn (\ref{3.27}) for these parameters is
shown in Figure (4). In the $(M_{c}, R_{c})$ domain with $K_{e} > 1$, one
measures the halo flux everywhere and this is provided entirely by extended
sources. In the domain with $K_{h} > 1$ and $K_{e} < 1$, one again measures the
halo flux everywhere but there are patches containing extended clusters where
the intensity is higher and given by eqn (\ref{3.16}). In both cases, there
will be Poisson fluctuations in the number of clusters along the line of sight
and this will give statistical anisotropies in addition to the angular
dependence given by eqn (\ref{3.22}). In the domain with $K_{h} < 1$, there is
no background, so one sees only individual clusters, the extended ones having
the intensity given by eqn (\ref{3.16}). Note that there will always be more
than one extended source in the satellite beam if
   \be
      M_{c} < 6 \times 10^{4} \; f_{h} \left( \frac{ \theta_{fov} }{ \mbox{3
      arcmin} } \right)^{2} \left( \frac{ R_{h} }{ \mbox{100 kpc} } \right) \;
      M_{\odot} \label{3.28}
   \ee
where $\theta_{fov}$ is the total field-of-view of the satellite (the pixel
field-of-view times the number of pixels).

\section{Cosmological background of Brown Dwarfs}

Although we have so far only focussed on the emission of BDs in our own
Galactic halo, there could also be a cosmological background from extragalactic
BDs. These could either be confined to other galactic halos or, if the BDs were
pregalactic in origin, they could be extragalactic. Karimabadi \& Blitz
\shortcite{karimabadi84} [hereafter KB] have calculated the intensity of the
background radiation from a critical density of BDs with a discrete IMF.
However, this seems an unlikely scenario since, in the standard model,
cosmological nucleosynthesis considerations require the BD density parameter to
satisfy
   \be
      \Omega_{bd}\leq \Omega_{B}\leq 0.015 h^{-2} \label{4.1}
   \ee
\cite{walker91}. Even unconventional models require $\Omega_{bd} < 0.09 \,
h^{-2}$ \cite{mathews93}, so background BDs could have at most a tenth of the
critical density. Note that, even if there are no intergalactic BDs (e.g.
because  the background dark matter is non-baryonic), one could still have
$\Omega_{bd} \approx 0.1$ in the form of halo BDs, so there would still be some
background flux from the emission of all the galaxies (see \S 4.3). In the
following analysis, we will leave $\Omega_{bd}$ unspecified but we assume that
the total density parameter is 1 even if $\Omega_{bd} < 1$. We also update the
KB results by using the luminosity and temperature scaling laws of Stevenson
(1986) and by allowing for an extended IMF. For the purposes of this section it
is irrelevant whether halo BDs are clumped into clusters, since such clusters
would be unresolved at cosmological distances.

The intensity of the BD background at present frequency $\nu$ in an
Einstein-de~Sitter universe is
   \be
      I_{\nu}(t_{0})=\rho_{bd}(t_{0})\frac{c}{4\pi}\int_{t_{f}}^{t_{0}}
      \Gamma_{\overline{\nu}}(t)\,dt \label{4.2}
   \ee
where $t_{0}$ is the present age of the universe, $\rho_{bd}(t_{0})$ is the
present BD density
   \be
      \rho_{bd}(t_{0})=1.88\times 10^{-29} \Omega_{bd}h^{2}\mbox{ g cm}^{-3}
      \label{4.3}
   \ee
$t_{f}$ is the formation time of the BDs (assuming that they all formed roughly
simultaneously) and $\Gamma_{\overline{\nu}}$ [given by eqn (\ref{3.15})] is
the specific luminosity per mass at the frequency $\overline{\nu}$, which, at
time $t$ corresponds to the present frequency $\nu$. In an
Einstein-de~Sitter universe, frequency evolves with time according to
   \be
      \overline{\nu} = \left(\frac{2}{3H_{0}t}\right)^{2/3} \nu \approx 0.75
      \, h^{-2/3}\tu^{-2/3} \nu \label{4.4}
   \ee
where $\tu \equiv t/10^{10}$ years. Hence the specific luminosity of an
individual BD at time $t$ is
   \begin{eqnarray}
      \lefteqn{ L_{\overline{\nu}}(m,t) \approx 1.19 \times 10^{15} \left[
      \frac{L(m,t)}{ L_{\odot}} \right] \left[ \frac{T(m,t)}{T_{\odot}}\right]
      ^{-4} } \nonumber \\ & & \;\;\;\;\;\;\;\;\;\;\;\;\;\;\;\;\;\;\;\;\;\;\;
      \;\;\;\;\;\;\;\;\;\;\;\;\;\;\;\; \times \left\{ \frac{ \fre^{3} h^{-2}
      \tu^{-2} } {\exp[0.062 \, \fre h^{-0.67} \tu^{-0.67} (T / T_{\odot})^{-1}
      ] -1} \right\} \mbox{ erg s$^{-1}$Hz}^{-1} \label{4.5}
   \end{eqnarray}
where the last term corresponds to a blackbody spectrum.

In order to evaluate $I_{\nu}(t_{0})$ accurately, the intensity contributions
from all of the evolutionary stages of BDs must be estimated. The major
evolutionary stages are: (1) the collapse of the protostellar cloud, in which
the luminosity increases until quasi-static equilibrium is reached; (2) the
Hayashi contraction phase, in which the luminosity decreases but the effective
temperature remains constant; (3) for BDs with masses larger than $0.015 \;
M_{\odot}$ we also have a deuterium-burning phase, lasting for around
$10^{7}-10^{8}$ years; (4) finally, all the BDs enter the long-lived degenerate
cooling phase. The intensity contribution from (1) is highly uncertain but
short-lived (around $10^{6}$ years) and so will be neglected here. We will also
omit the contribution from (3) since it is only relevant for higher mass BDs
and KB estimate that it would increase their calculated intensity by a factor
of at most 2. Since we are considering a BD mass range extending well below
$0.015 \; M_{\odot}$, the effect of deuterium-burning on the overall intensity
will be even less.

\subsection{The Hayashi contraction phase}

KB showed that BD luminosity during the contraction phase takes the form
   \be
      L\approx 2.9\times 10^{-6}(\mbd^{4.72}/\tbd^{2})^{1/3}\mbox{ }L_{\odot}
      \;\;\;\;\;(\tau_{col} < \tbd < \tau_{col}+\tau_{con}) \label{4.6}
   \ee
where $\tbd$ is the age of the BDs (measured from the time when the parent
clouds form), $\tau_{col}$ is the time taken for the protostellar collapse
phase (assumed to be about $10^{6}$ yr, so that $\tau_{col} \sim 10^{-4}$) and
$\tau_{con}$ is the duration of the Hayashi phase. The lower limit must be
specified since $L$ diverges at $\tbd = 0$. This equation was obtained from an
expression for the Kelvin-Helmholtz contraction timescale for very low mass
stars \cite{kumar63} and from an expression for the BD effective temperature
during this phase \cite{stevenson78}:
   \begin{eqnarray}
      t_{KH} & = & 4.98 \times 10^{5}\left(\frac{\mbd^{2}}{T_{3}^{4}}\right)
      \left(\frac{\ts R}{\ts R_{\odot}}\right)^{-3}\mbox{ yr} \label{4.7} \\
      T & \approx & 1960\,\mbd^{0.18}\mbox{ K} \label{4.8}
   \end{eqnarray}
where $T_{3}\equiv T/1000$ K. Eqn (\ref{4.6}) then comes from the blackbody
relation
   \be
      \frac{L}{L_{\odot}}=\left(\frac{R}{R_{\odot}}\right)^{2}\left(\frac{T}
      {T_{\odot}}\right)^{4} \label{4.9}
   \ee
where we have taken $T_{\odot}=5800$ K. Stevenson \shortcite{stevenson91} shows
that the radius for a fully degenerate BD is
   \be
      R\approx 0.15 \left( \frac{\mbd^{-0.25}}{1+0.57\mbd^{-0.5}} \right)^{4/3}
      R_{\odot} \label{4.10}
   \ee
Substituting eqn (\ref{4.10}) into (\ref{4.7}) gives the total time for the
contraction phase as
   \be
      \tau_{con}\approx 1.1\times 10^{-3} \mbd^{2.28}(1+0.57\mbd^{-0.5})^{4}
      \label{4.11}
   \ee
This increases from $10^{5}$ yr to $10^{9}$ yr as $m$ increases from $0.001 \;
M_{\odot}$ to $0.08 \; M_{\odot}$.

Eqns (\ref{4.6}), (\ref{4.8}) and (\ref{4.11}) allow us to calculate
$I_{\nu}(t_{0})|_{con}$, the intensity associated with the contraction phase.
Before doing so, we note that it is necessary to express the time dependence of
eqn (\ref{4.6}) in terms of the age of the universe $\tu$, rather than the age
of the BD $\tbd$. If we assume that BDs are all the same age, so that their
parent clouds collapse at the same time $\tau_{f}$, then $\tbd=\tu-\tau_{f}$
and eqn (\ref{4.6}) becomes
   \be
      L\approx 1.1\times 10^{28}\mbd^{1.57}(\tu-\tau_{f})^{-0.67}
      \mbox{ erg s}^{-1} \;\;\;\;\;\;\;\;\; (\tau_{f} + \tau_{col} < \tu <
      \tau_{f} + \tau_{col} + \tau_{con}) \label{4.12}
   \ee
{}From eqn (\ref{4.5}) we can therefore express $L_{\overline{\nu}}(m,t)$ as
   \be
      L_{\overline{\nu}}(t)|_{con}\approx 2.7\times 10^{11}\left[\frac{h^{-2}
      \tu^{-2}(\tu-\tau_{f})^{-0.67}\fre^{3}\mbd^{0.85}}{\exp(0.18h^{-0.67}
      \tu^{-0.67}\fre\mbd^{-0.18})-1}\right]\mbox{ erg s$^{-1}$Hz}
      ^{-1} \label{4.13}
   \ee
$I_{\nu}(t_{0})|_{con}$ can now be evaluated for specific BD IMFs by inserting
eqn (\ref{4.13}) into eqn (\ref{3.15}) for $\Gamma_{\overline{\nu}}$ and
integrating first over the time interval $\tau_{f} +
\tau_{col}$ to $\tau_{f} + \tau_{col} + \tau_{con}$, and then over the mass
range $m_{l}$ to $m_{u}$.

\subsection{The degenerate cooling phase}

For the degenerate cooling phase $(\tu > \tau_{f}+\tau_{col}+\tau_{con})$, we
can use eqns (\ref{2.6}) with $\tu-\tau_{f}$ in place of $\tbd$. Hence the
specific luminosity at time $t$ is
   \begin{eqnarray}
      \lefteqn{ L_{\overline{\nu}}(t)|_{deg}\approx 2\times 10^{13} \; h^{-2}
      \tu^{-2} \fre^{3} \mbd^{-0.66} } \;\;\;\;\;\;\;\;\;\;\;\;\;\;\;\;\;
      \nonumber \\  & & \times \{ \exp[ 1.63 \, h^{-0.67} \tu^{-0.67} (\tu-
      \tau_{f})^{0.94} \fre \mbd^{-0.79} \k^{-0.075} ]-1 \}^{-1} \mbox{ erg
      s$^{-1}$Hz}^{-1} \label{4.14}
   \end{eqnarray}
$I_{\nu}(t_{0})|_{deg}$ can now be evaluated in the same way as for the
Hayashi phase but with the integral in eqn (\ref{4.2}) running from
$\tau_{f}+ \tau_{col}+\tau_{con}$ to $\tau_{0}$ where $\tau_{0}\approx 0.65
h^{-1}$ for an Einstein-de~Sitter universe. Except for IMFs that are heavily
weighted towards the upper BD mass limit, the contraction phase contributes
most to the presently observed intensity for all reasonable values of $t_{f}$.
For discrete IMFs the ratio of the contributions of the degenerate and
contraction phases to the overall intensity is shown in Figure (6a). At any
particular frequency the ratio first increases and then decreases with
increasing $m$. It only exceeds 1 for low frequencies and low values of $m$; at
the \iras 12 $\mu$m frequency it is less than 0.2 for all $m$. For power law
IMFs with $m_{l} = 0.001 \; M_{\odot}$ and $m_{u} = 0.08 \; M_{\odot}$, the
ratio is shown in Figure (6b). At 12 $\mu$m it 0.04 for $x = 3$ and 0.1 for $x
= 0$ to 2. Both Figures assume $t_{f} = 10^{9}$ yr. For smaller values of
$t_{f}$ the contribution from the contraction phase is redshifted to lower
frequencies and so the ratio of the degenerate and contraction phase
contributions increases at high frequencies but decreases at low frequencies.

Figure (7a) shows the total expected intensity $I_{\nu}(t_{0})$ as a function
of frequency for the discrete IMF case with $\Omega_{bd}=1$ and $h=0.5$ for
different values of $m$. In the standard scenario, the maximum value of
$\Omega_{bd}$ is 0.06 for $h=0.5$, so the intensity could only be 0.06 that of
Figure (7a). Figure (7b) shows the expected intensity for power law IMFs with
$m_{l} = 0.001 \; M_{\odot}$ and $m_{u} = 0.08 \; M_{\odot}$. Curves (A) and
(B) correspond to the discrete IMF cases with $m = 0.001 \; M_{\odot}$ and $m =
0.08 \; M_{\odot}$ respectively. Comparison of Figures (7a) and (7b) shows that
each value of $x$ is well approximated by a particular value of $m$.

Note that the peak intensities of the halo and cosmological backgrounds are
comparable for $\Omega_{bd} = 0.1$ and $h = 0.5$ (peaking at around 500 Jy
sr$^{-1}$), although the halo background dominates at higher frequencies
provided $t_{f} < 1 Gyr$, since then the redshift effect dominates over the
contraction phase contribution for the cosmological background. The reason for
the two backgrounds having simiar intensities is that the cosmological
background scales as
   \be
      M_{u}^{} R_{u}^{-2} \approx \Omega_{bd} \left( \frac{  c H_{0} }{ 3 G }
      \right) \approx 0.5 \; \Omega_{bd} h \mbox{ g cm}^{-2} \label{4.15}
   \ee
where $M_{u}$ and $R_{u}$ are the mass and radius of the observable universe,
while the Galactic background scales as
   \be
      M_{h}^{} R_{h}^{-2} \approx 0.02 \left( \frac{ R_{h} }{ \mbox{100 kpc} }
      \right) \mbox{ g cm}^{-2} \label{4.16}
   \ee
Coincidentally, these values are very similar for $R_{h} = 100$ kpc,
$\Omega_{bd} = 0.1$ and $h = 0.5$. However, the Galactic background also
depends upon the Galactic coordinates, and this could be used to discriminate
it from a cosmological background.

\subsection{The background from halo brown dwarfs in other galaxies}

In this subsection we determine the circumstances in which the BDs in galactic
halos form a cosmological background. This will be the case providing the
covering factor of galactic halos exceeds 1. If the covering factor of BD halos
is less than unity, then one is merely looking for individual galaxies or
clusters of galaxies. This situation has been discussed by DM and will not be
dealt with here.

The differential covering factor of galaxies in the redshift interval $(z,z +
dz)$ is
   \be
      dK_{gal}(z) = dN_{gal}(z) \left[ \frac{ \Omega_{gal}(z) }{ 4 \pi }
\right]
      \label{4.17}
   \ee
where
   \be
      \Omega_{gal} = R_{h}^{2} \left( \frac{ H_{0} }{ c } \right)^{2}
      \left[  \frac{ (1 + z)^{3/2} }{ (1 + z)^{1/2} - 1 } \right]^{2}
      \label{4.18}
   \ee
is the solid angle subtended by a single galaxy with halo radius $R_{h}$ at
redshift $z$, and
   \be
      dN_{gal} = 16 \pi \; n_{gal} (c/H_{0})^{3} (1 + z)^{-5/2}
      [ (1 + z)^{1/2} - 1 ]^{2} \; dz \label{4.19}
   \ee
is the number of galaxies between $z$ and $z + dz$ (c.f. Kolb \& Turner 1990).
For simplicity we only consider galaxies like our own and assume these all have
the same halo radius. The present day number density of galaxies like our own
is $n_{gal} \simeq 1.5 \times 10^{-2} h^{3}$ Mpc$^{-3}$, so eqns (\ref{4.17})
to (\ref{4.19}) give
   \be
      \frac{ dK_{gal} }{ dz } = 5.7 \; h^{2} \left( \frac{ R_{h} }{ \mbox{ 100
      kpc} } \right)^{2} \left( \frac{ 1 + z }{ 10 } \right)^{1/2} \label{4.22}
   \ee
Hence the {\em total} coverage factor due to halos up to redshift $z$ is
   \be
      K_{gal} = \int_{0}^{z} \frac{ dK_{gal} }{ dz } \; dz  = 38 \; h^{2}
\left(
      \frac{ R_{h} }{ \mbox{ 100 kpc} } \right)^{2} \left( \frac{ 1 + z } { 10
      } \right)^{3/2} \label{4.23}
   \ee
For example, if $R_{h} = 50$ kpc and $h = 0.5$, then the coverage factor would
exceed 1 provided halos formed at redshifts greater than 5.6. If halo BDs are
Population III stars and therefore form at redshifts around 10, then $K_{gal}$
exceeds 1 for all values of $R_{h}$ and $h$. A more sophisticated calculation,
including the contribution to $K_{gal}$ from galaxies with a wider range of
masses is difficult since the dependence of $R_{h}$ upon galaxy mass is
unknown. In any case, this could only increase our estimate of $K_{gal}$.

\section{Current constraints and future prospects}

We now investigate the detectability of BDs in various scenarios. We first
examine the constraints imposed by the fact that, so far, analysis of {\em
IRAS}\/ data has not yielded any BD candidates. We then consider the prospects
that future infrared satellites such as \iso and \sirtf will either detect BDs
or place more stringent constraints on their contribution to the dark matter.

\subsection{\iras constraints}

The \iras faint source survey (FSS) covers 66\% of the sky $(|b| \geq
20^{\circ})$ over four wavelength bands: 12, 25, 60 and 100 $\mu$m. For BD
detection, the 12 $\mu$m band provides the only interesting constraint since,
at
longer wavelengths, the expected flux is well below the sensitivity of {\em
IRAS}\/. At 12 $\mu$m the FSS has a sensitivity of about 0.2 Jy and a pixel
field-of-view (FOV) of 15 arcsec \cite{moshir92}. In \S 2 we considered the
expected distance and flux of the {\em nearest}\/ BD. However, this calculation
only applies for observations made over the entire sky. For surveys covering
some fraction $\psi$ of the sky, the expected distance $\disdel$ to the nearest
BD {\em within the surveyed volume} is larger than that indicated by eqns
(\ref{2.17}) and eqn (\ref{2.22}) by a factor $\psi^{-1/3}$, so the expected
flux is decreased by a factor $\psi^{-2/3}$. Hence, for {\em IRAS}\/, one must
reduce the flux estimates by a factor of 1.32. Note that the probability that
$d$ is less than $\disdel$ is only 51\%. For 95\% confidence level constraints,
eqn (\ref{2.3}) shows that one must consider BDs at a distance $d = 1.61 \,
\disdel$, reducing the flux by another factor of 2.59.

Figures (1) and (2) show the expected flux of the nearest halo BD for all-sky
coverage. Even in the optimal case, in which $m = 0.07 \; M_{\odot}$, the
nearest BD would only have an expected flux of about 0.1 Jy at 12 $\mu$m which
is a factor of 2 below {\em IRAS}\/. This conclusion clearly extends to the
power law IMF case since the total flux must always lie within the optimum
discrete case. Therefore one cannot place any interesting constraints on the
density of unclustered halo BDs from \iras observations.

On the other hand, Figures (3a) and (3b) suggest that the disk scenario should
be strongly constrained by {\em IRAS}\/. Figure (3a) shows that, for an all-sky
survey, the expected flux from the nearest disk BD is 1.1 Jy at 12 $\mu$m for
the optimal case in which all the disk dark matter comprises $0.04 \;
M_{\odot}$ BDs. Therefore the 95\% confidence level (CL) lower limit on the
flux of such BDs within the \iras FSS is 0.32 Jy, and so the \iras FSS
sensitivity limit at 12 $\mu$m implies that $0.04 \; M_{\odot}$ BDs have $f_{d}
< 0.5$ at the 95\% CL and $f_{d} < 0.12$ at the 51\% CL. One could only have
$f_{d} = 1$ for $m < 0.02 \; M_{\odot}$ (95\% CL) or $m < 0.01 \; M_{\odot}$
(51\% CL). From Figure (3b) we see that, in the extended IMF case, all the disk
dark matter is in BDs only for $x > 2$  (51\% CL) or $x > 0$ (95\% CL) for
$m_{l} = 0.001 \; M_{\odot}$ and $m_{u} = 0.08 \; M_{\odot}$. Note that the
95\% CL limit is insensitive to $m_{l}$ but sensitive to $m_{u}$. As discussed
in \S 2.3, if the IMF of the disk BDs is a continuation of the more massive
stars, then it seems unlikely that disk BDs can be responsible for more than
about 10\% of the local disk dark matter. This constraint is an order of
magnitude stronger than that derived from the \iras FSS and implies that the
\iras survey will probably not yield any BDs.

For the cluster scenario, \iras already imposes constraints on the mass and
radius of clusters but these are only marginally interesting. We saw in \S 3
that the nearest clusters cannot be point sources without violating dynamical
constraints. In this case, for a discrete BD IMF, the \iras extended source
sensitivity of 1 MJy sr$^{-1}$ at 12 $\mu$m, along with eqn (\ref{3.13}),
requires
   \be
      \mc \ls 0.8 \, f_{\nu}(m)^{-1} \left( \frac{ R_{c} }{ \mbox{pc} } \right)
      ^{2} \label{5.1}
   \ee
where $f_{\nu}(m)$ is given by eqn (\ref{2.9}). This line is shown in Figure
(4) for the $m = 0.02 \; M_{\odot}$ case and the fact that it is so far to the
right of the ``Extended'' line confirms that the point source limit is
irrelevant. Comparison with the dynamical limits in Figure (4) shows that the
\iras extended source sensitivity limit is scarcely interesting: it only just
penetrates the dynamically permitted region even for a halo core radius of 8
kpc. For BDs larger than $0.02 \; M_{\odot}$, the \iras line moves to the right
but the evaporation constraint also moves up, so one still does not penetrate
the dynamically permitted region. In fact, $m = 0.02 \; M_{\odot}$ is the
optimum case. Figures (5a) and (5b) show that the \iras sensitivity is a factor
of 3 above the expected background at 12 $\mu$m if $M_{c} = 10^{6} \;
M_{\odot}$ and $R_{c} = 1$ pc.

\iras is also unable to place any constraints on the extended emission from a
halo or a cosmological background of BDs. Even a cosmological background with
the closure density would be two orders of magnitude too faint to have been
detected by {\em IRAS}, which is why the \iras limit is not shown in Figures
(7). Recall that, if the BDs are clustered, the background from our own halo is
only measured everywhere to the right of the line $K_{h} = 1$ in Figure (4) and
this line is well to the right of the \iras 12 $\mu$m line..

\subsection{2MASS constraints}

The 2 $\mu$m All-Sky Survey (2MASS) is a ground-based survey which aims to
search over the entire sky for sources at 1.2, 1.6 and 2.2 $\mu$m with a
limiting $10 \, \sigma$ K-band (2.2 $\mu$m) magnitude of 14 \cite{chester93}.
The survey should be underway in 1995, but Chester et al. have already
conducted a 1 deg$^{2}$ survey to test a prototype of the intended camera. In
the prototype survey they achieved a $10 \, \sigma$ limiting K-band magnitude
of 14.4 (corresponding to a $3 \, \sigma$ sensitivity of 0.35 mJy at 2.2
$\mu$m) over an area of about 0.8 deg$^{2}$ ($\psi = 1.6 \times 10^{-5}$).
The remaining 0.2 deg$^{2}$ was omitted from the analysis due to the globular
cluster M92 producing an excessively high source count in that region. For such
a sky coverage, the 2.2 $\mu$m sensitivity implies that $0.08 \; M_{\odot}$
unclustered disk BDs have $f_{d} < 0.16$ and only BDs with $m < 0.004 \;
M_{\odot}$ could have $f_{d} = 1$ (95\% CL). These constraints are more severe
than those from the \iras FSS, primarily because the 2.2 $\mu$m wavelength is
more favourable to the detection of disk BDs. One would expect the completed
survey to detect BDs with $m > 0.01 \; M_{\odot}$ if they comprise all the disk
dark matter (95\% CL). The prototype survey does not, however, constrain the
contribution of unclustered BDs to the halo dark matter. One requires the
survey to sweep at least 8 deg$^{2}$ of sky before one can expect to detect
even the most massive halo BDs. However, the completed survey should permit the
detection of halo BDs with $m > 0.03 \; M_{\odot}$.

\subsection{Future observations}

Before the end of the decade two major satellite observatories will be
launched. The first will be \iso (Infrared Space Observatory), whose {\em
ISOCAM}\/ instrument will consist of a detector array of $32 \times 32$ pixels.
In the 6 arcsec pixel FOV mode, this gives a detector FOV of 3 arcmin with a
point source sensitivity of 50 $\mu$Jy and an extended source sensitivity of $3
\times 10^{4}$ Jy sr$^{-1}$ \cite{cesarsky92}. These figures are based on
observations centred at 6.75 $\mu$m with a $10 \, \sigma$ detection limit in an
integration time of 3600 s. The second satellite will be \sirtf (Space Infrared
Telescope Facility). The wide field instrument on board \sirtf will have a
detector FOV of 7 arcmin. It will consist of up to $256 \times 256$ pixels and
have an estimated $10 \, \sigma$ point source sensitivity of 15 $\mu$Jy for an
integration time of 100 s per field at 3 $\mu$m \cite{werner92}. In this
section we will base constraints on $3 \, \sigma$ sensitivities for integration
times of 100 s. For an integration time $t_{pf}$ the 6.75 $\mu$m \iso $y \,
\sigma$ point source sensitivity is
   \be
      F_{\nu}^{sat}(\mbox{{\em ISO}\/}) = 90 \; \mu \mbox{Jy} \; \left( \frac{
      y }{ 3 } \right) \left( \frac{ t_{pf} }{ \mbox{100 s} } \right)^{-1/2}
      \;\;\;\;\;\;\;\; (\lambda = 6.75 \; \mu \mbox{m}) \label{5.2}
   \ee
The equivalent 3 $\mu$m \sirtf point source sensitivity is 4.5 $\mu$Jy. Another
relevant factor is the satellite FOV. The FOV corresponds to a sky coverage of
$\psi_{sat} = 6.1 \times 10^{-8}$ per integration for \iso and $\psi_{sat} =
3.3 \times 10^{-7}$ for {\em SIRTF}\/. The larger the FOV, the greater the area
of sky covered in a single exposure and therefore the shorter the total
observing time required to survey a given area. Unlike {\em IRAS}\/, \iso and
\sirtf are observatory-class rather than survey instruments, which means that
they will be making pointed observations over selected areas of the sky rather
than conducting all-sky surveys . However, as in \S 5.1, we can obtain
estimates for the BD flux based on the amount of sky surveyed.

Figures (1) and (2) show the \iso point source sensitivity for a total
observation time $t_{obs} = 10$ days (excluding the time taken between
exposures) and an integration time per field of 100 s, giving a total of 8640
surveyed fields or $\psi = 5.3 \times 10^{-4}$. The \sirtf point source
sensitivity is not shown but is an order of magnitude stronger. Eqns
(\ref{2.8}) and (\ref{2.18}) show that the expected flux is proportional to
$f^{2/3}$ where $f$ stands for $f_{h}$ in the halo case and $f_{d}$ in
the disk case, so if the satellite detects a BD with a flux $F_{sat}$ we can
infer that the most likely value of $f$ is
   \be
      f = \left( \frac{ F_{sat} }{ F_{exp}^{bd} } \right)^{3/2} \label{5.3}
   \ee
where $F_{sat}$ is the limiting total flux of the observing satellite
integrated over the bandpass $(\nu_{1}, \nu_{2})$ of the observations and
$F_{exp}^{bd}$ is the total BD flux expected if $f = 1$ integrated over the
same bandpass:
   \be
      F_{sat} = \int_{\nu_{1}}^{\nu_{2}} F_{\nu}^{sat} \; d\nu \;\;\;\;\;\;\;\;
      F_{exp}^{bd} = \int_{\nu_{1}}^{\nu_{2}} F_{\nu}^{exp} \; d\nu \label{5.4}
   \ee
Eqns (\ref{2.18}) and (\ref{2.23}) imply
   \begin{eqnarray}
      F_{\nu}^{bd} & \approx & 0.1 \; \psi^{2/3} g(m)^{2/3}
      \mbd^{1/3} f_{\nu} (m,\tau,\kappa) \mbox{ Jy} \;\;\;\;\;\;\;\;
      \mbox{(halo)} \label{5.5} \\
      F_{\nu}^{bd} & \approx & 0.8 \; \psi^{2/3} g(m)^{2/3}
      \mbd^{1/3} f_{\nu} (m,\tau,\kappa) \mbox{ Jy} \;\;\;\;\;\;\;\;
      \mbox{(disk)} \label{5.6}
   \end{eqnarray}
It should be stressed that eqn (\ref{5.3}) only determines $f$ as a function of
$m$. Without some independent measure of the BD temperature, $m$ itself is
unknown. Therefore observations in a single band would not specify the scenario
uniquely. If the satellite makes no detection, then eqn (\ref{5.3}) merely
gives a 51\% CL upper limit on $f$. The 95\% CL upper limits would be smaller
by a factor of 4.

Figure (8) shows the $3 \, \sigma$ constraints on $f_{h}$ and $f_{d}$ for a
range of discrete BD IMFs in the event of \iso not detecting any BDs after a
total observation time of 1, 10 and 100 days with integration times of 100 s.
For power law BD IMFs, constraints can be obtained on a particular mass range
$(m , \alpha m)$ by dividing $f_{h} \, (f_{d})$ in Figure (8) by the quantity
$g(m)$ defined in eqn (\ref{2.13}). Since one does not know $g(m)$ {\em a
priori}, eqn (\ref{5.3}) really only specifies the combination $f g(m)$. The
time required to detect a BD would then be larger than indicated in Figure (8)
by a factor of $g(m)^{-1}$. It can be seen that observation times longer that
10 days would be required to detect halo BDs below $0.03 \; M_{\odot}$ or disk
BDs below $0.006 \; M_{\odot}$.

Figure (4) show that clusters of BDs should also be detectable by \iso for a
small range of cluster parameters. The gain in angular resolution of \iso over
\iras means that BD clusters will be extended at even larger distances: since
$\theta_{res} = 6$ arcsec for {\em ISO}\/, eqn (\ref{3.6}) indicates that the
critical distance is $D = 7 \times 10^{4} R_{c}$, which means that {\em every}
cluster in the halo should be an extended source. Figures (5a) and (5b) show
that the expected intensity of clusters with mass $10^{6} M_{\odot}$ and radius
1 pc at 6.75 $\mu$m. In the optimal case, in which clusters BDs with a mass of
$0.08 M_{\odot}$, \iso should be able to reach the required intensity level
with only a few seconds integration time per field. For other cluster masses
and radii the integration time required scales as $M R_{c}^{-3}$. The number of
fields which need to be searched depends upon the FOV of the satellite
$\theta_{fov}$ and the covering factor $K_{e}$ of the extended clusters.
The satellite must survey a fraction of sky exceeding $K_{e}$ in order
to constrain the cluster scenario. From eqn (\ref{3.28}), if no detection were
made after searching $N$ fields, one could infer a constraint
   \be
      f_{h} < 0.17 \; \mc \left( \frac{ \psi_{sat} }{ 6.1 \times 10^{-8} }
      \right)^{-1} \left( \frac{ R_{h} }{ \mbox{100 kpc} } \right)^{-1} \left(
      \frac{ N }{ 100 } \right)^{-1}  \label{5.8}
   \ee
For integration times of 100 s, \iso should be sensitive to intensities as low
as $5 \times 10^{4}$ Jy sr$^{-1}$ at 6.75 $\mu$m $(3 \, \sigma)$, which is
enough to detect clusters comprising BDs with a discrete IMF if $m \geq 0.015
\;
M_{\odot}$ or with a power law IMF if $x \ls 2.5$ (for $m_{l} = 0.001 \;
M_{\odot}$ and $m_{u} = 0.08 \; M_{\odot}$). Hence, for a total observation
time exceeding 1.9 days with integration times of 100 s (giving $N \geq 1640$),
\iso should either detect $10^{6} \; M_{\odot}$ clusters with $R_{c} = 1$ pc
and $m \geq 0.015 \; M_{\odot}$ or $x \ls 2.5$ or else constrain their
contribution to $f_{h} < 0.01$.

Detection of halo background emission will require longer total observation
times. For observations in the direction of the Galactic pole, $J (b = \pi/2, a
= 2\mbox{ kpc}) \approx 1.4$ and so eqn (\ref{3.17}) implies $I_{\nu}^{halo}
\ls 300 \; f_{h}$ Jy sr$^{-1}$ for $m < 0.08 M_{\odot}$ and  $\tbd = \k = 1$.
This upper limit is weaker than that of Adams \& Walker \shortcite{adams90} by
a factor of $6-7$, primarily due to the different scaling laws used for $L$ and
$T$; they adopt ``case D'' of Burrows, Hubbard \& Lunine \shortcite{burrows89}.
Such a low intensity means that observation times exceeding 50 days are
required to detect any halo emission. (The results of Adams \& Walker give an
observation time of 1 day, which reflects how uncertainties in BD models affect
our estimates.) As discussed earlier, cosmological emission from intergalactic
BDs or BDs in galactic halos would be of a similar intensity and so would
require observations of comparable duration. Figures (7a) and (7b) show that a
negative result from \iso would imply $m \leq 0.03 \; M_{\odot}$ for a discrete
IMF and $x > 1.5$ for a power law IMF if $\Omega_{bd} = 1$.

\subsection{Conclusions}

The preceeding discussion shows that the present \iras constraints are
uninteresting for unclustered BDs. The sensitivity of the \iras survey falls
short of the expected flux levels of halo BDs in the Solar neighbourhood
even in the optimum case with $m = 0.07 \; M_{\odot}$. Although disk BDs could
have been detected by \iras if they comprised all of the disk dark matter, a
simple extrapolation of the disk IMF suggests that this is unlikely. The
cluster scenario is also poorly constrained by \iras, stronger constraints on
the permitted mass and radius of clusters coming from dynamical considerations.
The \iras constraints only overlap the dynamical constraints for a large halo
core radius and a very narrow range of cluster parameters. The extended source
sensitivity of \iras also falls well short of the sensitivity required to
detect any Galactic or cosmological background. Thus BD clusters are still
viable candidates for the halo dark matter and they may also be massive enough
to explain the observed Galactic disk heating.

The next generation of infrared satellites will have significant gains in
sensitivity over \iras and should be able to detect the nearest halo BDs, be
they clustered or unclustered, down to very low masses. The timescale for \iso
to detect unclustered halo BDs is of order hours for $0.08 \; M_{\odot}$, days
for $0.04 \; M_{\odot}$ and months for $0.02 \; M_{\odot}$. For disk BDs the
timescale is of order hours for $0.01 \; M_{\odot}$. If the BDs are clustered,
their detectability depends crucially on the mass and radius of the clusters
but \iso should certainly be able to probe the parameter range permitted by
dynamical constraints. If the clusters cover the sky, then one is just looking
for a Galactic background. However, background emission from BDs may be
difficult to detect even with \iso or {\em SIRTF}\/. Although an $\Omega_{bd} =
1$ cosmological background would be detectable, a more plausible $\Omega_{bd}
\sim 0.1$ background (e.g. due to galactic halos) would be very difficult to
detect, as would the background emission from our own Galaxy. Only if \iso
failed to detect a background after 50 days, would serious constraints be
imposed on the BD scenario and one might then be forced to invoke non-baryonic
candidates as a solution to the Galactic dark matter problem. However, at
present BDs remain excellent candidates, especially since nucleosynthesis
calculations suggest that the dark baryon density must exceed the visible
baryon density by at least a factor of two.

The main problem in calculating the detectability of BDs and BD backgrounds is
that the present systematic uncertainties in the expected luminosities and
temperatures of BDs translate into very large uncertainties in the constraints
or, equivalently, in the calculated observation times required for verifying a
particular scenario. These uncertainties are in part due to uncertainties in
opacity determinations and they may be reduced by the time \iso or \sirtf are
in
operation. As we have used the lowest estimates of the opacity in our
calculations, refinements in these calculations should reduce the estimated
observing times.

Although we have confined attention to the detection of BDs, one can clearly
use the same arguments to constrain M-dwarfs. Indeed the \iras constraints in
Figures (1) and (2) already exclude an appreciable fraction of the dark matter
being in objects bigger than $0.08 \; M_{\odot}$. This confirms the conclusion
of earlier work, based on searches for extended emission around other galaxies.
We should stress that, for the halo case, there is no reason to assume that the
BD masses should extend all the way up to $0.08 \; M_{\odot}$ and, as can be
seen in Figure (2c), this is crucial as regards to the halo BD detectability.

A particularly interesting source of a cosmological background would be a
population of dwarf galaxies mostly composed of BDs. Since the dark to visible
mass ratio may increase as one goes to smaller galaxies \cite{persic90}, it is
possible that most of the dark baryons are in this form. Dwarf galaxies
may be regarded as BD clusters, so the analysis of \S 3 (which was confined to
BD clusters in our own Galaxy) may be adapted to cover this scenario. It is
easy to check that the nearest dwarf galaxy would not be detectable but the
cosmological background from all such galaxies could be for sufficiently long
observation times. In this scenario, the background associated with dwarf
galaxies would be easier to detect than that from normal galaxies provided they
had a larger density.

We have stressed that there are four strategies for seeking the infrared
emission from BDs: (1) looking for the nearest discrete source, be it a single
BD or a cluster of BDs; (2) looking for our own halo emission; (3) looking for
the halo emission from other galaxies or clusters of galaxies; (4) searching
for a cosmological background. It is therefore important to compare the
efficacies of these different methods. Although method (1) requires the least
sensitivity, it has the disadvantage that one must search a significant area of
sky. If the BDs are clustered, the sensitivity required is reduced but one
needs to search a larger area. Methods (2) and (4) have the advantage that one
can look for a longer time; one can hardly search for many integration times in
a particular patch of sky in the hope of finding a discrete source. Method (3)
was discussed by DM. Coincidentally, the cosmological background is likely to
be comparable to the Galactic background, so methods (2) and (4) both require
comparable sensitivity.

\large\section*{ACKNOWLEDGEMENTS}

The authors would like to thank Tom Chester, Simon Green, Michael
Rowan-Robinson and John Watson for helpful discussions. The authors also thank
the anonymous referee for helpful comments on an earlier version of this paper.
EJK is supported by an SERC studentship.

\appendix
\section{The discrete representation of an extended IMF}

In this appendix we calculate the errors involved in representing a continuous
IMF by a finite number of strips. From eqns (\ref{2.12}) and (\ref{2.18}), the
error involved in taking the mass to be $m$ across the strip $(m,\alpha m)$ is
less than 50\% provided
   \be
      0.5 < \varepsilon = \frac{ g(\alpha m)^{2/3} (\alpha m)^{1.04} }{
      g(m)^{2/3} m^{1.04} } \approx \alpha^{(7-2x)/3} < 1.5 \label{a1}
   \ee
Table \ref{tab1} shows the ranges of $\alpha$ that satisfy eqn (\ref{a1}) for
various values of $x$. Note that it is the lower limit on $\varepsilon$ in eqn
(\ref{a1}) which is relevant for $x > 3.5$ since the exponent of $\alpha$ is
then negative. Also shown in the Table is the value of $\varepsilon$
associated with the choices of $\alpha$ used in Figure (2a), (2b) and (2c).
These are $\alpha = 1.55$, $\alpha = 1.23$ and $\alpha = 1.26$, respectively.
We see that eqn (\ref{a1}) is always satisfied for these choices if $x>1$.

We must also choose $\alpha$ to be large enough that the error introduced by
neglecting the contribution to the flux from the neighbouring mass strips is
kept small. The frequency at which the flux from the mass range $(m, \alpha m)$
peaks is given by eqn (\ref{2.10}) and the corresponding flux is given by eqn
(\ref{2.12}) multiplied by $g(m)^{2/3}$. The contribution to the flux at this
frequency from the mass range $(\alpha^{-1} m, m)$ is
   \be
      F_{\nu(m)}^{bd} (\alpha^{-1}m) \simeq 0.3 \; \alpha^{2x/3}
      \mbd^{2(3.56-x)/3} \chi [\exp(2.8 \alpha^{0.79})-1]^{-1} \mbox{ Jy}
      \label{a2}
   \ee
where we have taken $\tbd = \k = 1$ and
   \be
      \chi = \left[ \left( \frac{ m_{l} }{ 0.01 \; M_{\odot} } \right)^{x-2}
      \left( \frac{ \alpha^{2-x} - 1 }{ \beta^{2-x} - 1 } \right) \right]^{2/3}
      \label{a3}
   \ee
with $\beta \equiv m_{u} / m_{l}$. The contribution to the flux from the mass
range $(\alpha m, \alpha^{2} m)$ is
   \be
      F_{\nu(m)}^{bd} (\alpha m) \simeq 0.3 \; \alpha^{-2x/3}
      \mbd^{2(3.56-x)/3} \chi [\exp(2.8 \alpha^{-0.79})-1]^{-1} \mbox{ Jy}
      \label{a4}
   \ee
Eqns (\ref{2.12}), (\ref{a2}) and (\ref{a4}) therefore imply that the ratio of
the sum of the fluxes from the neighbouring strips to that from $(m, \alpha m)$
is
   \be
      \varepsilon = 17 \, \alpha^{2x/3} [\exp(2.8 \alpha^{0.79}) - 1]^{-1} +
      17 \, \alpha^{-2x/3} [\exp(2.8 \alpha^{-0.79}) - 1]^{-1} \label{a5}
   \ee
\begin{table}
 \centering
\caption{Errors introduced in attributing the flux from the mass range
$(m,\alpha m)$ to a single mass $m$. The second column shows the upper limit on
$\alpha$ required to keep this below 50\% for various values of $x$. The values
of $\varepsilon$ associated with the choices of $\alpha$ in Figures (2a), (2b)
and (2c) are shown in columns 3, 4 and 5 respectively.}
 \label{tab1}
 \begin{tabular}{lcccc}
  $x$ & $\alpha_{max}$ & $\varepsilon (\alpha = 1.55)$ &
     $\varepsilon (\alpha = 1.23)$ & $\varepsilon (\alpha = 1.26)$ \\
  0 & 1.2 & 2.80 & 1.60 & 1.70 \\
  1 & 1.3 & 2.05 & 1.40 & 1.48 \\
  2 & 1.5 & 1.55 & 1.23 & 1.26 \\
  3 & 3.4 & 1.16 & 1.07 & 1.08 \\
  4 & 8.0 & 1.14 & 1.07 & 1.08 \\
 \end{tabular}
\end{table}
Table \ref{tab2} gives the minimum value of $\alpha$ required for $\varepsilon
< 2$ and the value of $\varepsilon$ for the values of $\alpha$ used in Figures
(2). It shows that $\varepsilon$ is fairly insensitive to $\alpha$ and $x$ in
the relevant ranges and it is 2 or less for $x > 1$.

\begin{table}
   \centering
\caption{Errors arising from neglecting the flux contribution of neighbouring
mass ranges at the frequency of peak flux for the mass range $(m, \alpha m)$.
The second column gives the lower limit on $\alpha$ required to keep this below
2. The final three columns tabulate the value of $\varepsilon$ for the choices
of $\alpha$ used in Figures (2a), (2b) and (2c) respectively.}
   \label{tab2}
   \begin{tabular}{lcccc}
$x$ & $\alpha_{min}$ & $\varepsilon (\alpha = 1.55)$ & $\varepsilon (\alpha =
   1.23)$ & $\varepsilon (\alpha = 1.26)$ \\
0 & -- & 2.94 & 2.29 & 2.34 \\
1 & -- & 2.38 & 2.18 & 2.19 \\
2 & 1.9 & 2.03 & 2.09 & 2.09 \\
3 & 1.4 & 1.85 & 2.05 & 2.03 \\
4 & 1.4 & 1.83 & 2.05 & 2.03 \\
   \end{tabular}
\end{table}

The errors are worst for $x = 0$ and 1 since, for these values, the flux at
{\em all} frequencies is dominated by the {\em largest} masses [i.e. by the
mass range $(\alpha^{-1}m_{u}, m_{u})$]. As explained in the text, this is
expected for $x > 1.2$. In these cases the envelopes in Figures (2) are
equivalent to the discrete case with $m = \alpha^{-1} m_{u}$ multiplied by the
factor $g(m)^{2/3}$. This is evident by the absence of ``kinks'' in the
envelopes for $x = 0$ and 1. Since the second term in eqn (\ref{a5}) is absent
when one uses the strip $(\alpha^{-1} m_{u}, m_{u})$, the errors in Table
\ref{tab2} are not relevant for $x = 0$ and 1.

Comparison of Tables \ref{tab1} and \ref{tab2} shows that the choices of
$\alpha$ used in Figures (2) give errors of about a factor of 2 in the flux
estimates of eqn (\ref{2.18}). Changing the values of $\alpha$ would reduce one
source of error but increase the other. Thus our choices of $\alpha$ are a
reasonable compromise. We are also constrained to choose values of $\alpha$
that give an integer number of strips between $m_{l}$ and $m_{u}$, so in
general one cannot use the optimum value. Note that a factor of 2 is anyway
comparable to the uncertainty in some of the physical parameters of BDs (such
as the effect of grain opacity on the effective temperature of BDs). The
analysis used in \S 2.2 and \S2.3 will always underestimate rather than
overestimate the flux at a given frequency and so any constraints on the mass
function inferred from observations will be on the conservative side.

\parindent=0cm
\parskip=9pt
\large\section*{FIGURE CAPTIONS}

{\bf Figure (1):} The expected flux from the nearest halo BD for various
discrete IMFs. The halo is assumed to comprise BDs with a local density of
$0.01
\; M_{\odot}$ pc$^{-3}$, an age of 10 Gyr and an opacity of 0.01
cm$^{2}$g$^{-1}$. The \iras point source sensitivity at 12 $\mu$m is shown
and this is a factor of two above the predicted flux even for BDs with the
optimal mass of $0.07 \; M_{\odot}$. This assumes an all-sky survey; since
the \iras faint source survey does not cover the entire sky, the constraints on
the density of BDs are really a factor 1.32 weaker. The expected $3 \, \sigma$
\iso 6.75 $\mu$m sensitivity is also shown, assuming a total observation time
of 10 days with an integration time of 100 s.

{\bf Figures (2):} The expected flux from the nearest halo BD for various power
law IMFs, assuming a BD age of 10 Gyr, an opacity of 0.01 cm$^{2}$g$^{-1}$ and
a local density of $0.01 \; M_{\odot}$ pc$^{-3}$. Figure (2a) corresponds to
$m_{u} = 0.08 \; M_{\odot}$ and $m_{l} = 0.001 \; M_{\odot}$; Figure (2b) to
$m_{u} = 0.08 \; M_{\odot}$ and $m_{l} = 0.01 \; M_{\odot}$; and Figure (2c) to
$m_{u} = 0.01 \; M_{\odot}$ and $m_{l} = 0.001 \; M_{\odot}$. The solid lines
represent the discrete mass cases for $m = m_{u}$ (A) and $m = m_{l}$ (B). The
dashed lines correspond to power law IMFs with $x$ in the range 0 to 4. Since
different mass ranges dominate at different frequencies, the curves are the
envelopes of the fluxes from BDs of different masses. In order to determine
these curves, the IMF is divided into 10 logarithmically equal mass intervals
between $m_{l}$ and $m_{u}$. ``Kinks'' in the curves for $x \geq 2$ are due to
the discontinuities this introduces in the IMF. The \iras and \iso point source
sensitivities are shown for the same parameters as in Figure (1).

{\bf Figures (3):} The expected flux from the nearest disk BD. Figure (3a)
corresponds to various discrete IMFs. Figure (3b) corresponds to power law IMFs
with $m_{l} = 0.001 \; M_{\odot}$ and $m_{u} = 0.08 \; M_{\odot}$; (A) and (B)
show the discrete IMF cases corresponding to $m = m_{u}$ and $m = m_{l}$
respectively. We have assumed a disk BD density of $0.15 \; M_{\odot}$
pc$^{-3}$, a BD age of 5 Gyr and an opacity of 1 cm$^{2}$g$^{-1}$. The \iras
and \iso point source sensitivities assume the same parameters as in Figure
(1).

{\bf Figure (4):} \iras and \iso constraints on the cluster scenario together
with the dynamical constraints of Lacey \& Ostriker \shortcite{lacey85},
Carr \& Lacey \shortcite{carr87} and Carr \& Sakellariadou \shortcite{carr93}.
The dynamical constraints (discussed in the text and indicated by the bold
lines) are sensitive to the age of the Galaxy $t_{g}$ and the halo radius
$R_{h}$. The dynamical friction limits also depend on the halo core radius $a$
and are shown for $a = 2$ kpc and $a = 8$ kpc. The evaporation limit is
sensitive to the individual BD mass and is shown for a mass of $0.02 \;
M_{\odot}$. To the right of the dotted line labelled ``Extended'' the nearest
cluster appears to \iras as an extended source and the permitted cluster regime
lies well within this region. The \iras 12 $\mu$m (solid line) and \iso 6.75
$\mu$m (dot-dash line) extended source sensitivities depend on the individual
BD mass and the best constraints are obtained for clusters comprising $0.02 \;
M_{\odot}$ BDs. The $3 \, \sigma$ \iso sensitivity assumes a total observation
time of 1.1 hrs (assuming $R_{h} = 100$ kpc) and an integration time of 100 s.
The covering factor of all the clusters in our halo in the direction of the
Galactic anticentre exceeds 1 to the right of the $K_{h} = 1$ lines and, in
this region, one observes the Galactic background everywhere. To the left of
this line, one observes individual clusters and emission is only expected over
certain patches of the sky.

{\bf Figures (5):} The expected intensity of a cluster of mass $10^{6}
M_{\odot}$ and radius 1 pc which is close enough to be extended, assuming a BD
age of 10 Gyr and opacity of 0.01 cm$^{2}$g$^{-1}$. The nearest clusters are
always extended and it is possible that all of the clusters in the Galaxy are.
The expected intensity for various discrete BD IMFs are plotted in Figure (5a).
In Figure (5b) the dashed lines correspond to power law IMFs with $m_{l} =
0.001 \; M_{\odot}$, $m_{u} = 0.08 \; M_{\odot}$ and $x$ ranging from 0 to 4,
while (A) and (B) correspond to the limiting discrete IMF cases. For the values
of $M_{c}$ and $R_{c}$ shown, the \iras extended source sensitivity at 12
$\mu$m is a factor of two greater than the predicted intensity even in the
optimum case. However, for other cluster masses or radii, the intensity scales
as $M_{c} R_{c}^{-2}$, so \iras can still provide constraints on clusters with
sufficiently small radii. The \iso $3 \, \sigma$ extended source sensitivity at
6.75 $\mu$m is also shown for a total observation time of 400 s and an
integration time of 100 s. If the covering factor of the clusters exceeds 1,
there will be a general Galactic background with an intensity below that shown
in the figure by a factor 3600.

{\bf Figures (6):} The ratio of the contribution to the observed cosmological
background from the BD degenerate cooling and contraction phases. Figure (6a)
shows the ratio for various discrete IMF. For high masses the contraction phase
dominates at all frequencies but it is {\em most} dominant at low frequencies.
For low masses the degenerate cooling phase is unimportant for all but the
lowest frequencies, where it actually {\em dominates}. Figure (6b) shows the
ratio for various power law mass spectra with $m_{l} = 0.001 \; M_{\odot}$ and
$m_{u} = 0.08 \; M_{\odot}$. For high values of $x$ the degenerate cooling
phase dominates at the lowest frequencies, is unimportant at the intermediate
frequencies but then contributes significantly as one goes to higher
frequencies. The contraction phase always dominates for low values of $x$. Both
Figures assume that BDs form at a cosmic time of 1 Gyr and that the Hubble
constant is 50 km s$^{-1}$Mpc$^{-1}$. For earlier formation times the
contribution of the contraction phase is redshifted to lower frequencies and so
becomes less important at high frequencies.

{\bf Figures (7):} The expected intensity of the cosmological background
generated by the contraction and degenerate cooling phases of a critical
density of BDs. Figure (7a) shows the expected intensity for various discrete
BD IMFs, whilst Figure (7b) shows the maximum (A) and minimum (B) discrete
cases, along with power law spectra with $m_{l} = 0.001 \; M_{\odot}$, $m_{u} =
0.08 \; M_{\odot}$ and $x$ in the range 0 to 4. We assume a BD opacity of 0.01
cm$^{2}$g$^{-1}$, a Hubble parameter of $h = 0.5$, and a BD formation time of 1
Gyr. Such a scenario is constrained by {\em ISO}\/ but not {\em IRAS}\/; the
\iso $3 \, \sigma$ 6.75 $\mu$m limit is shown for a total observation time of
10 hours and an integration times of 100 s. The background intensity for values
of $\Omega_{bd}$ compatible with nucleosynthesis calculations (i.e. $0.04 \leq
\Omega_{bd} \leq 0.06$ for a Hubble parameter of 0.5) would be 15 times less
and would be hard to detect even with {\em SIRTF}\/.

{\bf Figure (8):} Constraints on the BD halo mass fraction $f_{h}$ (solid line)
and disk mass fraction $f_{d}$ (broken line) as a function of BD mass for
discrete IMFs implied by \iso not detecting BDs after a total observation time
$t_{obs}$ of 1, 10 and 100 days and an integration time $t_{pf}$ of 100 s in
the $5 - 8.5 \; \mu$m bandpass. To find the constraints on some mass range $(m,
\alpha m)$ for power law IMFs, one must multiply $f_{h}$ or $f_{d}$ by the
factor $g(m)$ given by eqn (\ref{2.13}). We assume a BD age of 10 Gyr and an
opacity of 0.01 cm$^{2}$g$^{-1}$. For the same $t_{obs}$ and $t_{pf}$, \sirtf
should either detect halo BDs or provide constraints an order of magnitude
stronger.

\end{document}